\documentclass[aip,jcp, floatfix,preprint,citeautoscript]{revtex4-1}
\setcitestyle{super}

\usepackage{graphicx}
\usepackage{subfigure}
\usepackage{setspace}[1.5] 
\usepackage{amsmath, amsthm, amssymb}
\usepackage{xfrac}
\usepackage{dcolumn}
\usepackage{color}
\usepackage[flushleft]{threeparttable}

\newcommand{\etal}{\textit{et al.}}

\newcommand{\ccsdt}{$\Delta$CCSD(T)}
\newcommand{\aza}{1,2-azaborine}

\begin{document}

\date{\today}

\title{Water on BN doped benzene: A hard test for
  exchange-correlation functionals and the impact of exact exchange on
  weak binding}

\author{Yasmine S. Al-Hamdani}
\affiliation{Thomas Young Centre and London Centre for Nanotechnology,
  17--19 Gordon Street, London, WC1H 0AH, U.K.}
\affiliation{Department of Chemistry, University College London, 20
  Gordon Street, London, WC1H 0AJ, U.K.}

\author{Dario Alf\`{e}}
\affiliation{Thomas Young Centre and London Centre for Nanotechnology,
  17--19 Gordon Street, London, WC1H 0AH, U.K.}
\affiliation{Department of Earth Sciences, University College London,
  Gower Street, London WC1E 6BT, U.K.}

\author{O. Anatole von Lilienfeld}
\affiliation{Institute of Physical Chemistry, Department of Chemistry,
  University of Basel, Klingelbergstrasse 80 CH-4056 Basel,
  Switzerland}
\affiliation{Argonne National Laboratories, 9700 S. Cass
  Avenue Argonne, Illinois 60439, USA}

\author{Angelos Michaelides}
\email{angelos.michaelides@ucl.ac.uk}
\affiliation{Thomas Young Centre and London Centre for Nanotechnology,
  17--19 Gordon Street, London, WC1H 0AH, U.K.}
\affiliation{Department of Chemistry, University College London, 20
  Gordon Street, London, WC1H 0AJ, U.K.}

\begin{abstract}
Density functional theory (DFT) studies of weakly interacting
complexes have recently focused on the importance of van der Waals
dispersion forces whereas, the role of exchange has received far less
attention. Here, by exploiting the subtle binding between water and a
boron and nitrogen doped benzene derivative (\aza) we show how exact
exchange can alter the binding conformation within a complex.
Benchmark values have been calculated for three orientations of the
water monomer on \aza\ from explicitly correlated quantum chemical
methods, and we have also used diffusion quantum Monte Carlo. For a
host of popular DFT exchange-correlation functionals we show that the
lack of exact exchange leads to the wrong lowest energy orientation of
water on \aza.  As such, we suggest that a high proportion of exact
exchange and the associated improvement in the electronic structure
could be needed for the accurate prediction of physisorption sites on
doped surfaces and in complex organic molecules. Meanwhile to predict
correct absolute interaction energies an accurate description of
exchange needs to be augmented by dispersion inclusive functionals,
and certain non-local van der Waals functionals (optB88- and
optB86b-vdW) perform very well for absolute interaction
energies. Through a comparison with water on benzene and borazine
(B$_3$N$_3$H$_6$) we show that these results could have implications
for the interaction of water with doped graphene surfaces, and suggest
a possible way of tuning the interaction energy.
\end{abstract}
\maketitle

\section{Introduction}
An accurate description of the structures and energies of weakly
interacting systems is important in materials science and biology, but
it is often difficult to obtain reference data either theoretically or
experimentally. A key challenge lies in the ability to capture small
energy differences -- on the order of a few meV -- that can have
drastic effects on the structure. For example, water has several
distinct ice polymorphs that have lattice energies within $35$
meV/H$_2$O of each
other\cite{Whalley1984,Galli2012,Santra2011,Santra2013,Gillan2013,Slater2014}.
Likewise for water clusters, most notably the water hexamer, there are
several isomers that have energies within just 5
meV/H$_2$O\cite{Santra2008,Gillan2013,Pedulla1998}. Furthermore, in
biological applications there can be numerous shallow energy minima
with different conformations. Particularly for complex organic
systems, predicting the exact lowest energy conformation is crucial to
determine the crystal structure of drugs\cite{Price2009,Price2014} and
the mechanisms by which proteins function\cite{Timasheff1993}.

One particularly interesting weakly interacting system of relevance to
potential applications in clean energy, water purification, hydrogen
storage, and bio-sensing \cite{Siria,Lei2013,bn_exp3,graph2,graph3},
is the interaction of water with layered materials. Notably
interfacial water on graphene, hexagonal boron nitride (h-BN) and
hybrid composites of these materials. Indeed thanks to remarkable
advances in combining boron, nitrogen and carbon atoms in a cyclic
aromatic arrangement it is now possible to create 2-dimensional sheets
with carefully structured regions of carbon and boron
nitride\cite{hybBNG,hybBNG1,synBNDG}. Despite the growing number of
studies for water on h-BN\cite{bn_exp,bn_exp4,Marti2011} and
graphene\cite{graph2,graph3,graph4,graph5,Jenness2010,Hamada2012,alfe2}
there are no direct measurements of adsorption energies for the water
monomer, and the theoretical adsorption energies for these systems
vary significantly across different high accuracy
methods\cite{wg_bind1,alfe2,Jenness2010,Rubes2009}.

One can use smaller model systems for
graphene\cite{Feller2000,Xu2005,Sudiarta2006,Rubes2009,Jenness2010}
and h-BN, such as benzene and the inorganic counterpart borazine
(B$_3$N$_3$H$_6$), to help understand the interaction with water. With
these small molecules, it is possible to use high accuracy methods to
calculate benchmark interaction energies and binding
conformations,\cite{wb_bind2,wb_bind3,wb_bind1,alfe1,Wu2012} that
would otherwise be infeasible for the extended surfaces. Given the
shortage of reference data across these systems, there is a strong
incentive to deliver accurate benchmark calculations, and for our
purposes, we require a model system that is a hybrid of benzene and
borazine. With that in mind, we study the weak binding between water
and an aromatic molecule known as 1,2-dihydro-1,2-azaborine (or
\aza\ for short) as a reference system. The \aza\ molecule shares many
similarities to benzene, with the clear distinction being the
asymmetry of the molecule due to boron and nitrogen substitution (see
Fig. \ref{3structures}). The mixture of boron, nitrogen and carbon
atoms in this molecule makes it a suitable model for benchmarking in
relation to extended surfaces that are hybrids of graphene and
h-BN\cite{homoBenz}. Moreover, the asymmetry makes \aza\ an ideal
system for testing the performance of computational methods because
each atom in the ring has a distinct chemical environment that serves
as a tag, in contrast to benzene in which carbon atoms are obviously
indistinguishable.

Previous work on the water-benzene interaction and other weak
interaction systems has demonstrated that methods without long-range
correlation fail to account for dispersion interactions that are
important for the interaction energy
\cite{wg_bind1,alfe1,alfe2,Bjorkman2012,Bjorkman2014,Antony2006,Graziano2012,Carrasco2013}.
However, most studies of such systems have focused on the description
of long-range correlation, and fewer have shown that the underlying
exchange approximation can also have an impact on the binding
interaction\cite{Langreth2006,exchangeDFT,biswajit}. Here, we have
examined the water/\aza\ system with coupled cluster with single,
double and perturbative triple excitations (CCSD(T)), and a range of
density functional theory\cite{HK,KS} (DFT) exchange-correlation (xc)
functionals. We find that many exchange-correlation functionals
predict the wrong conformation of water on \aza, and this problem is
solved by including a high proportion of exact exchange, highlighting
the need for improvements in existing models of
exchange\cite{vdwpers,cohen,Becke2014,Kieron2012}. As part of this
study we have also tested diffusion quantum Monte Carlo (DMC). For
selected systems DMC has been shown to produce very accurate results
\cite{alfe1,alfe2,Mattsson,Jordan2013} including ``subchemical''
accuracy ($<1$ kcal mol$^{-1}$) in dispersion dominated
systems\cite{Mitas}. Here we show that DMC does well with respect to
the CCSD(T) benchmarks, again achieving subchemical accuracy.

This short article will start with a brief summary of the employed
methods, followed by results from a set of calculations that allow us
to directly compare the performance of different xc functionals
with other explicitly correlated methods. After analysis and
discussion of the various DFT results on the intermediate
\aza\ system, we investigate the effect of the boron nitride doping,
by also studying the interaction of water with the pure systems of
benzene and borazine. We close with a discussion and some general
conclusions.

\section{Methods}\label{METHOD}

\subsection{Computational Methods}

We have undertaken a series of quantum chemical calculations using
Dunning's augmented correlation consistent basis sets
(aug-cc-pVXZ)\cite{dunning1,dunning2,dunning3}. Second order
M{\o}ller-Plesset perturbation theory (MP2) with up to aug-cc-pV5Z
basis sets \footnote{We have also investigated the magnitude of basis
  set superposition error by applying Boys and Bernardi's counterpoise
  correction\protect\cite{cpBB}, but the correction was not included
  in the CBS extrapolation (see SI\cite{SI_ref} for more details).}
along with CCSD(T) calculations at the aug-cc-pVTZ level have been
conducted. Due to the unfavorable scaling of CCSD(T), it is more
feasible to conduct MP2 calculations with larger basis sets, deducing
the complete basis set (CBS) limit, and subsequently calculating the
$\Delta$CCSD(T) value of absolute interaction energy at the CBS
limit. For a description of this procedure along with analysis of
errors, the reader is referred to the recent work of Sherrill and
coworkers\cite{Sherrill}. Regarding the CBS limit, various
extrapolation schemes have been discussed
\cite{cbs1,cbs2a,cbs2b,cbs3,cbs4} and we have chosen to use the one
proposed by Halkier \etal\cite{cbs2a,cbs2b,cbs3} Gaussian03 \cite{g03}
was used for the Hartree-Fock (HF) and post-HF calculations.

The initial single particle wavefunctions for use in DMC were obtained
from DFT plane-wave (PW) calculations using the PWSCF package
\cite{pwscf} and Trail and Needs pseudopotentials (PPs) were used for
all atoms in the system\cite{TN1,TN2}, warranting a standard 300 Ry
energy cut-off. Previous work by Ma \etal\cite{alfe1} indicates that
weak binding energies are not overly sensitive to the trial
wavefunctions (TWs), having tested a few xc functionals (including
hybrids) and also HF. We have generated TWs using the local density
approximation (LDA)\cite{LDA} and also the Perdew-Burke-Ernzerhof
(PBE)\cite{PBE} xc functional. The resulting wavefunctions were
expanded in terms of B-splines \cite{bsplines} for efficiency. DMC
calculations have been performed using the CASINO code\cite{casino},
and we have used Slater-Jastrow type TWs, in which the Jastrow factor
contains electron-nucleus, electron-electron, and
electron-electron-nucleus terms. We used a combination of DMC
calculations using 16000 walkers across 160 cores and 64000 walkers
across 640 cores. Final DMC results have been derived by the weighted
averaging of the results and errors. Time steps of 0.0025, 0.005 and
0.01 a.u. have been tested and the locality approximation was
utilized\cite{locapp}. We obtained statistical error bars for our
interaction energies of $\pm3$ meV, which corresponds to $1\sigma$.

VASP 5.3.2 \cite{vasp1,vasp2,vasp3,vasp4} was used for all the DFT
calculations. VASP employs plane-wave basis sets and uses projector
augmented wave (PAW) potentials\cite{PAW,paw2} to model the core
region of atoms. After a series of convergence tests for the
plane-wave cut-off energy and unit cell, we chose to use a 500 eV
cut-off energy and a 15 \AA\ length cubic unit cell, along with
$\Gamma$-point sampling of reciprocal space.

There is, of course, an almost endless list of xc functionals that
could be considered, and here we benchmark a selection of fairly
widely used functionals. The functionals tested include PBE \cite{PBE}
which is a generalized gradient approximation (GGA) functional that
does not contain long-range correlation. We have also considered the
hybrid xc functionals which contain a proportion of exact exchange:
PBE0 \cite{PBE0a,PBE0b} and
B3LYP\cite{b3lypA,b3lypB,b3lypC,b3lypD}. There have been many
developments to include van der Waals (vdW) dispersion in xc
functionals, as discussed in the perspective of Klime{\v{s}} and
Michaelides \cite{vdwpers} and reviewed by Grimme\cite{vdwreview}, and
here we have tested several of these vdW-inclusive DFT
approaches. Specifically, PBE-D2\cite{D2}, a semi-empirical functional
that contains Grimme's D2 correction, and also two correction schemes
from Tkatchenko and Scheffler, namely vdW-TS\cite{TS} and
vdW-TS+SCS\cite{MBD}, referred to here as TS and TS+SCS,
respectively. Using the TS and TS+SCS schemes, C$_6$ coefficients and
vdW radii are determined from ground state electron
densities\cite{TS}, whilst TS+SCS also includes long-range screening
effects\cite{MBD}. The TS and TS+SCS corrections will be applied to
PBE, PBE0 and B3LYP\footnote{As the TS and TS+SCS schemes are
  implemented in the later versions of VASP, VASP.5.3.3 was used for
  these particular calculations.}. We have also tested several
non-local vdW functionals from the ``vdW-DF''
family\cite{vdwDF,vdwfuncs,vdwimp,vdwDF2,Cooper2010,Hyldgaard2014,Hamada2014,Voorhis2009,Voorhis2010,Gironcoli2013,Jacobsen2012,BjorkmanPRB2012}.
The vdW-DFs considered include the original vdW-DF which we refer to
as revPBE-vdW\cite{vdwDF}, several optimized vdW functionals
(optPBE-vdW\cite{vdwfuncs}, optB88-vdW\cite{vdwfuncs}, and
optB86b-vdW\cite{vdwimp}), and also vdW-DF2\cite{vdwDF2}. The exchange
part of revPBE-vdW and the optimized vdW functionals are different but
they all share the same non-local correlation part. In vdW-DF2, both
the exchange and correlation components have been modified.

\subsection{Water and \aza\ Setup}\label{benchmark}

The absolute interaction energy, $E_{int}$, that we refer to
throughout this study is defined as,
\begin{equation}
E_{int}=E^{tot}_{com}-E^{tot}_{sub}-E^{tot}_{wat}\label{BE}
\end{equation}
where $E^{tot}_{com}$ is the total energy of the bound complex between
water and the substrate (\aza\, benzene or borazine), and
$E^{tot}_{sub}$ and $E^{tot}_{wat}$ are the total energies of relaxed
substrate molecule and water, respectively. 

\begin{figure}[ht]
\centering \includegraphics[width=1\textwidth]{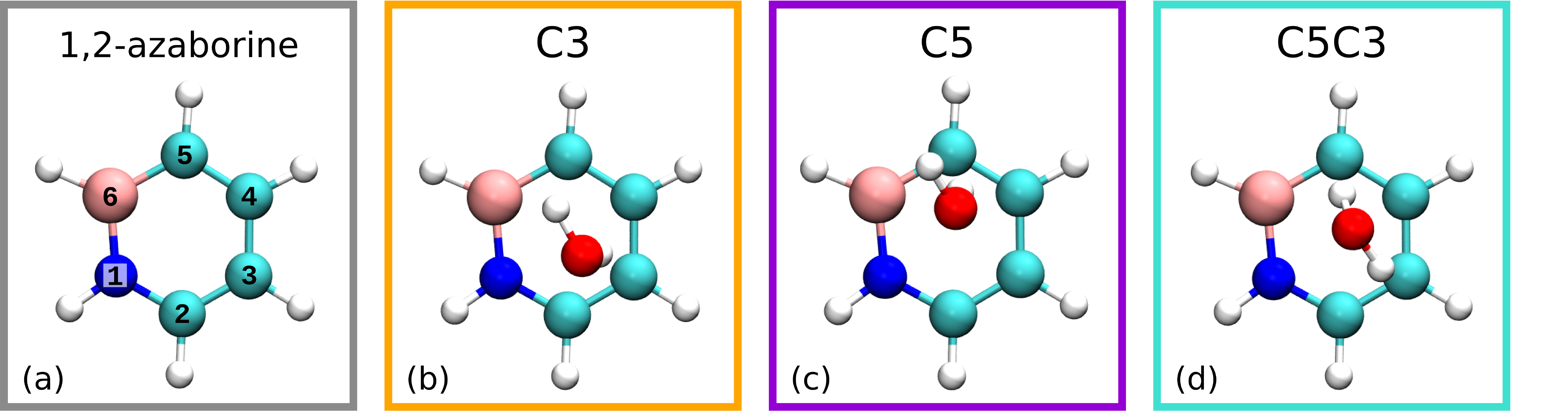}
\caption{(a) \aza\ molecule: 
  the carbon atoms are numbered according to their positions, nitrogen
  being 1 and boron being 6. The three distinct binding configurations
  of water (oxygen in red and hydrogen in white) on \aza\ are: C3 (b),
  C5 (c), and C5C3 (d).}\label{3structures}
\end{figure}

\pagebreak
In \aza\ the electronic environment of individual carbon atoms differs
due to the asymmetry introduced through the boron and nitrogen
atoms. We use the numbering scheme shown in Fig. \ref{3structures}(a)
to make the distinction between the atoms. 

We obtained three distinct orientations of water over \aza\ using
PBE-D2\footnote{A 10 \AA\ long cubic cell was used with standard PBE
  PAW potentials and a 500 eV cut-off energy. Convergence criteria of
  $10^{-6}$ eV for the wavefunction optimization and $0.01$
  eV/\AA\ for the forces were used.} in order to carry out the
benchmarking study, namely C3, C5 and C5C3, also depicted in
Fig. \ref{3structures}. The nomenclature of the complexes refers to
the specific carbon atoms involved in hydrogen bonding, keeping in
line with the numbering scheme in Fig. \ref{3structures}(a). The
hydrogen atoms of water point toward carbon-3 and carbon-5 in all
three complexes; most likely as a result of the higher electron charge
around these carbon atoms due to the conjugation of the localized lone
pair of electrons from the nitrogen atom. Indeed, Bader analysis shows
that carbon-3 and carbon-5 have a larger atomic volume compared to
carbon-2 and carbon-4, as a result of increased electron charge around
them. Some characteristic structural parameters are listed in Table
\ref{structparams} for the three different configurations. Although
the PBE-D2 geometries are not benchmark accuracy, there have been
various studies on polymer crystals and layered
materials\cite{Ramprasad2012,Angyan2010,Zaoui2012}, and even for water
on graphene\cite{wg_bind1}, indicating that PBE-D2 can provide
reasonable structures for weakly interacting systems. Interested
readers may also refer to the Supporting Information\cite{SI_ref} (SI)
for bond lengths as well as the structures of fully relaxed complexes
with MP2 and several of the xc functionals that we test.
\begin{table}[ht]
\caption{\label{structparams} Perpendicular separation distances (in
  {\AA}ngstrom) of the oxygen atom to the plane of the 1,2-azaborine
  ring (R$_\text{{O-plane}}$), and hydrogen atoms of water to C3 and
  C5 (R$_\text{HW-C3}$ and R$_\text{HW-C5}$). The shortest C-H
  distance is reported in each case.}
\begin{ruledtabular}
\begin{tabular}{cccc}
 & C3 & C5 & C5C3\\\hline
R$_\text{O-plane}$ & $3.24$ & $3.22$ & $3.15$ \\
R$_\text{HW-C3}$   & $2.44$ & $4.42$ & $2.75$ \\
R$_\text{HW-C5}$   & $3.50$ & $2.39$ & $2.51$ \\
\end{tabular}
\end{ruledtabular}
\end{table}

\section{Results and Discussion}\label{results}
\subsection{Stability Trends for water on 1,2-azaborine}\label{resone}

We computed benchmark absolute interaction energies for the three
water adsorption complexes (C3, C5, and C5C3) mentioned in Section
\ref{benchmark}, using \ccsdt. The results for \ccsdt\ have been
extrapolated to the CBS limit and the computed interaction energies
reveal that C3 is the most stable binding configuration with an
interaction energy of $-155$ meV, followed by C5 ($-146$ meV) and C5C3
($-143$ meV), as listed in Table \ref{benchmarks}.

The interaction energies we computed with DMC are in reasonable
agreement with \ccsdt; within 5--14 meV depending on the adsorption
structure considered and the TW used. This level of agreement is in
line with several recent studies in which DMC has been compared to
coupled cluster\cite{alfe1,alfe2,Mitas,Benali2014}. The small
difference between DMC and \ccsdt\ interaction energies could be due
to issues such as the use of PPs in DMC, the fixed node approximation
in DMC, or the approximations used to obtain the \ccsdt\ values
(including CBS extrapolation)\cite{Sherrill}.  With regard to the
relative stabilities of the complexes, the DMC results suggest the
same trend as \ccsdt: C3 is more stable than C5 and C5 is more stable
than C5C3. Considering the statistical error bars for each DMC
interaction energy however, C3 and C5 could be degenerate according to
DMC with LDA TWs. Whilst with PBE TWs, C3 is 9 meV more stable than
C5, indicating a better trend prediction with the latter. Note the
total energies using LDA TWs are slightly lower than those obtained
with PBE TWs\footnote{The LDA TWs give rise to total energies that are
  only $\sim20$--$30$ meV lower than total energies obtained from PBE
  TWs, whereby the total energies are in the region of $\sim1500$ eV.}
and since DMC is a variational method, we consider the interaction
energies with LDA TWs to be slightly more reliable in this particular
system. Therefore, once again, it appears that DMC is useful for
obtaining reliable interaction energies, but there is also an inherent
difficulty in using a stochastic method like DMC, to clearly
distinguish between complexes with very small energy differences.

\begin{table}[ht]
\caption{\label{benchmarks}Absolute interaction energies in (meV) of
  water on 1,2-azaborine, using the structures shown in
  Fig. \ref{3structures}. The benchmark values from \ccsdt\ are
  presented in addition to DFT, HF, MP2 and DMC results. Lowest
  energies for each method are highlighted in bold.}
\begin{ruledtabular}
\begin{tabular}{@{}lccc@{}}
Methods     & C3               & C5     &  C5C3       \\ \hline 
PBE         & $ -98$    &$\mathbf{-110}$   & $-87 $   \\\hline
PBE-D2      & $-188$    &$\mathbf{-196}$   & $-195$   \\
PBE+TS      & $-168$    &$\mathbf{-174}$   & $-169$   \\
PBE+TS+SCS  & $-162$    &$\mathbf{-169}$   & $-161$   \\\hline
revPBE-vdW  & $-115$    &$\mathbf{-127}$   & $ -96$   \\
optPBE-vdW  & $-159$    &$\mathbf{-170}$   & $-148$   \\
optB88-vdW  & $-154$    &$\mathbf{-164}$   & $-148$   \\
optB86b-vdW & $-157$    &$\mathbf{-167}$   & $-150$   \\
vdW-DF2     & $-134$    &$\mathbf{-143}$   & $-122$   \\\hline
PBE0$^{0.25}$ & $-105$  &$\mathbf{-110}$   & $-92$    \\
PBE0$^{0.50}$ & $\mathbf{-114}$ & $-112$   & $-100$   \\
PBE0$^{0.75}$ & $\mathbf{-124}$ & $-116$   & $-109$   \\\hline
PBE0$^{0.25}$+TS &$\mathbf{-174}$ & $-173$ & $\mathbf{-174}$      \\
PBE0$^{0.25}$+TS+SCS & $\mathbf{-168}$ & $\mathbf{-168}$ & $-165$ \\
PBE0$^{0.75}$+TS & $\mathbf{-191}$     & $-177$          & $-190$ \\
PBE0$^{0.75}$+TS+SCS & $\mathbf{-181}$ & $-168$          & $-178$ \\\hline
B3LYP$^{0.20}$& $-65$   & $\mathbf{-73}$   & $-46$    \\
B3LYP$^{0.40}$& $-95$   & $\mathbf{-97}$   & $-78$    \\
B3LYP$^{0.60}$& $\mathbf{-125}$ & $-121$   & $-110$   \\\hline
B3LYP$^{0.20}$+TS+SCS & $-128$  & $\mathbf{-131}$ & $-120$ \\
B3LYP$^{0.60}$+TS+SCS & $\mathbf{-187}$  & $-178$   &$-183$ \\\hline
HF            & $ \mathbf{-22}$ & $  -8$   & $ +15$   \\
MP2/CBS       & $\mathbf{-164}$ & $-157$   & $-152$   \\
DMC ($\Psi_{\textrm{LDA}}$)   & $\mathbf{-144 \pm 2}$  &$-141 \pm 2$ & $-132 \pm 3$  \\
DMC ($\Psi_{\textrm{PBE}}$)   & $\mathbf{-143 \pm 3}$  &$-134 \pm 3$ & $-129 \pm 3$  \\\hline
\ccsdt/CBS    & $\mathbf{-155}$        &$-146$       & $-143$        \\ 
\end{tabular}
\end{ruledtabular}
\end{table}
\begin{figure}[ht]
\centering
\includegraphics[width=1\textwidth]{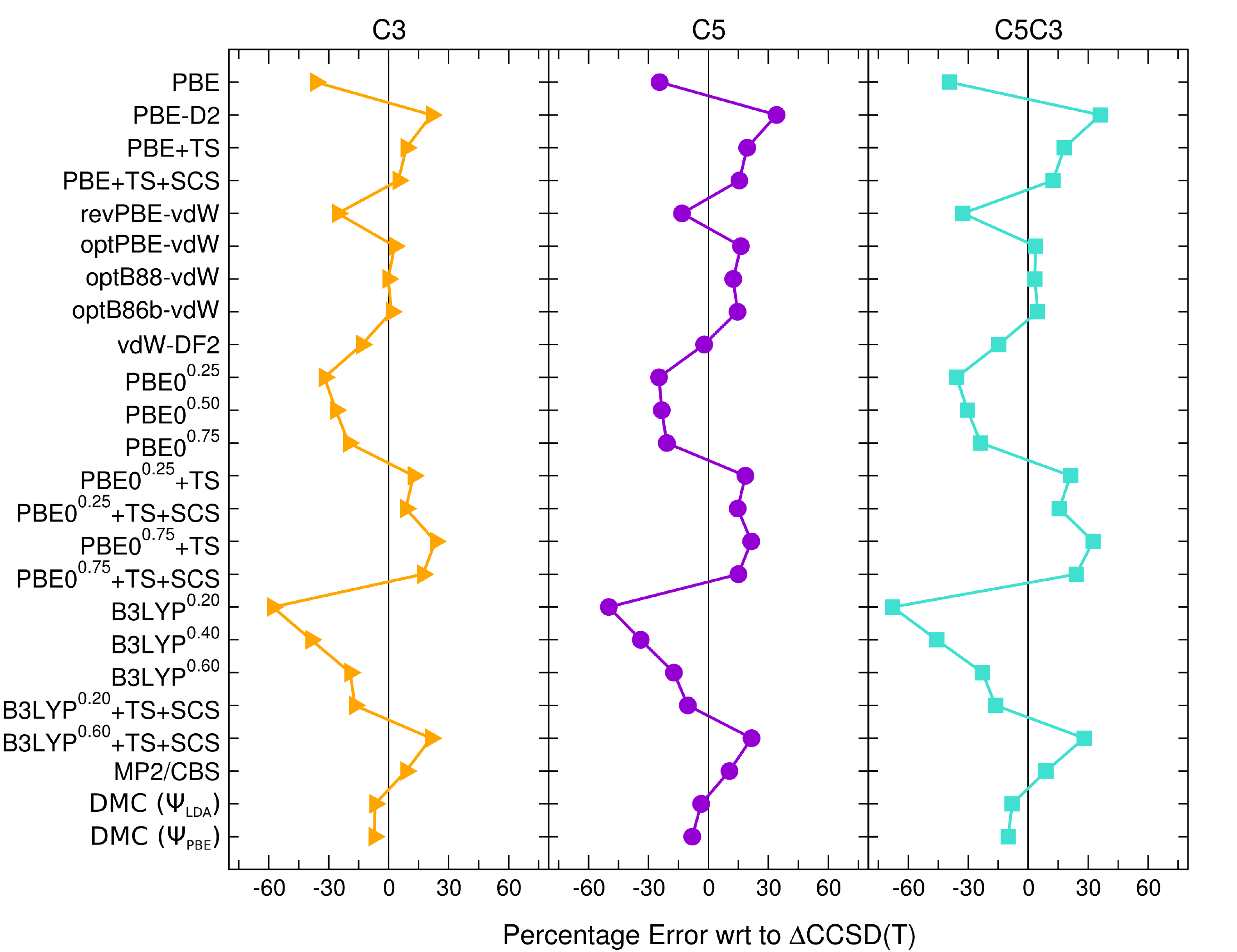}
\caption{The percentage difference between \ccsdt\ interaction energy
  and that from DMC, MP2 and the various DFT xc functionals, of water
  to 1,2-azaborine. The solid black lines at zero represent the
  \ccsdt\ reference. Note that DMC results with both LDA
  ($\Psi_{\textrm{LDA}}$) and PBE ($\Psi_{\textrm{PBE}}$) TWs are
  reported and that the HF results are off the chart. The superscripts
  for the hybrid PBE0 and B3LYP functionals indicate different
  proportions of exact exchange.}\label{qcdft}
\end{figure}

If we now consider the results obtained from the various DFT xc
functionals, we find that the trend obtained is in stark contrast to
the benchmark \ccsdt\ results; with GGA and dispersion inclusive xc
functionals showing preference for C5 instead of C3. However, before
discussing this in detail, we analyze the performance of the xc
functionals in terms of absolute interaction energies, and compare to
\ccsdt\ as illustrated in Fig. \ref{qcdft}. For the most stable C3
complex the best agreement with \ccsdt\ is given by optB88-vdW,
optB86b-vdW, and optPBE-vdW: remarkably less than $3\%$ errors. The
other vdW functionals do not perform as well, with vdW-DF2
underbinding by $15\%$ and revPBE-vdW underbinding by $25\%$. As
anticipated, PBE is strongly underbinding by almost $35\%$ due to the
lack of long-range correlation, whilst dispersion corrected PBE-D2
overestimates the binding by $20\%$. The TS and TS+SCS corrections
perform significantly better than the D2 correction, with only $10\%$
and $5\%$ errors, respectively.

One can see from Fig. \ref{qcdft} that the percentage error lines
across C3, C5, and C5C3 have a very similar form, but they shift with
regard to the reference \ccsdt\ binding energy. This means that for
the C5 complex vdW-DF2 is providing the best agreement with
\ccsdt\ ($2\%$ error) and the optimized vdW functionals are
overbinding by $10$--$15\%$. Whereas for C5C3, the xc functionals
perform in a similar manner as for C3, with the optimized vdW
functionals performing the best once again ($<5\%$ error). Of PBE and
its dispersion corrected forms, PBE-TS+SCS performs the best for all
three complexes (underbinding by $5$-$15\%$).  Note that MP2
consistently overbinds all three structures by $\sim9\%$.

Regardless of the absolute interaction energies of PBE, dispersion
corrected PBE and the non-local vdW functionals, they all fail to
predict C3 as the most stable complex. In addition, the TS and TS+SCS
corrections are not satisfactory as they stabilize the C5C3 complex
such that it becomes degenerate with either C3 or C5.  Clearly with a
fairly flat potential energy surface, the difference between C3 and C5
is a considerable challenge for the xc functionals.  According to the
benchmark \ccsdt\ values, C5 is only 9 meV less stable than C3, making
it difficult to assign the source of error that leads to so many
different xc functionals predicting the wrong trend.

One possible source of error is an inadequate description of exchange
and to address this we initially performed HF calculations. We find
that with HF the trend is correctly predicted with C3 as the most
stable configuration, despite the lack of correlation and
highly underestimated interaction energies. The HF results suggest
that the lack of exact exchange is perhaps the main reason for many of
the xc functionals predicting C5 instead of C3. For further insight,
HF symmetry adapted perturbation theory (HF-SAPT) calculations
\footnote{HF-SAPT calculations were performed using Molpro
  2010\protect\cite{MOLPRO_brief} and an aug-cc-pVDZ basis set for the
  C3 and C5 complexes.}  revealed that it is the exchange-repulsion
energy (mostly electrostatic) that puts the binding energy in favor
of C3 and not C5.

\begin{figure}[ht]
\centering \includegraphics[width=0.75\textwidth]{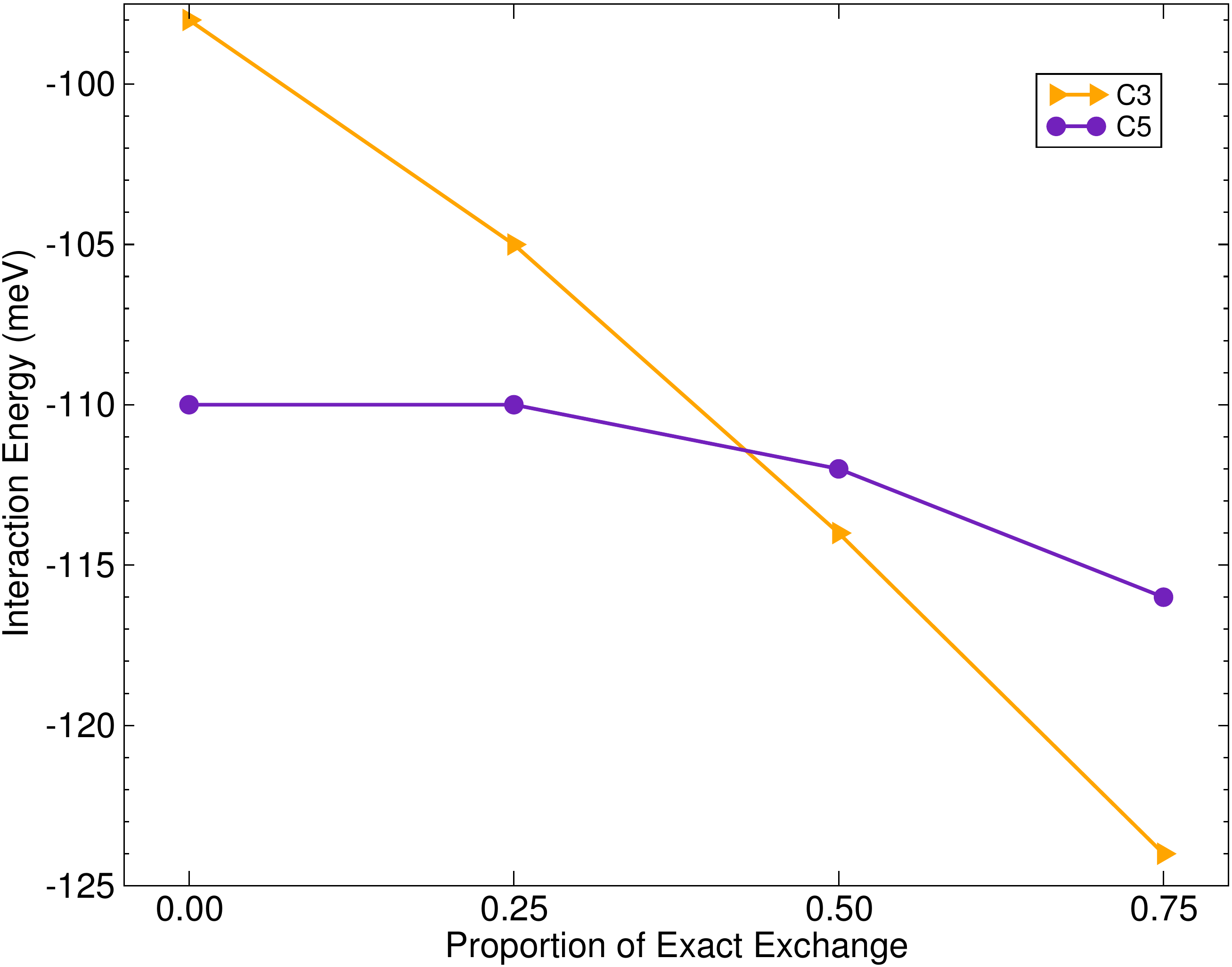}
\caption{The interaction energy of the C3 and C5 complexes are plotted
  against the proportion of exact exchange in PBE0-style calculations,
  with zero exact exchange corresponding to PBE.}\label{IEEXX}
\end{figure}

Guided by this insight, we computed binding energies using the hybrid
PBE0 and B3LYP functionals with varying amounts of exact exchange. The
results are listed in Table \ref{benchmarks} with the proportion of
exact exchange indicated by the superscripts, along with hybrid
functional results corrected with the TS and TS+SCS vdW schemes. In
addition, Fig. \ref{IEEXX} shows how the interaction energies of the
C3 and C5 complexes vary with the proportion of exact exchange with
PBE0-like functionals.
Using standard PBE0$^{0.25}$ the lowest energy complex is C5, but as
the percentage of exact exchange is increased to $50\%$ and above, C3
becomes the most stable complex. Similarly with B3LYP, $60\%$ exact
exchange is needed in order to switch the site preference to
C3. However PBE0$^{0.75}$ and B3LYP$^{0.60}$ still underbind by $20\%$
with respect to \ccsdt, and including the TS+SCS corrections leads to
overbinding by the same amount. The combination of the hybrid
functionals with TS and TS+SCS corrections tends to decrease the
energy differences between the different configurations. Depending on
the particular combination of exact exchange and TS-like dispersion,
both C3 and C5 can be degenerate or almost degenerate or even all
three structures can be almost degenerate. Overall we learn from these
various calculations that a high proportion of exact exchange improves
the relative energies of the various structures. However, still, in
the future a more refined description of long-range correlation and
exchange is needed in order to predict the correct trend and absolute
interaction energies.

A change in the stability trend of weakly interacting complexes due to
the amount of exact exchange is not limited to the systems studied
here; Thonhauser \etal\cite{Langreth2006} observed a similar
distortion in the ordering of conformers in their study of
monosubstituted benzene dimers, which they corrected by using HF
exchange. The need for a very high fraction of exact exchange in xc
functionals to give correct predictions warrants further discussion.
We have very carefully looked at the electronic structures obtained
from the various functionals for the different complexes and a
thorough inspection of the individual energy contributions to the
absolute interaction energies was particularly informative. By
decomposing the interaction energies into the kinetic, potential,
Hartree, exchange and correlation energies, we have analyzed the
effect of increased exact exchange on the individual energy
contributions for C3 and C5. We find that the main distinguishing
feature between C3 and C5, is that the kinetic energy and the
repulsive terms (Hartree and exchange energies) change to different
extents as the fraction of exact exchange is increased. Kinetic energy
favors C5 but with a high fraction of exact exchange, the C3 complex
is stabilized by less repulsion (Hartree and exchange) than in
C5. This observation is in accord with the HF-SAPT results mentioned
earlier.

It is known that exact exchange localizes the electrons and alleviates
the surplus delocalization which is otherwise present in standard GGA
calculations. The localization of electrons determines the electron
density distribution over a system and therefore determines the
interaction sites within complexes. In addition to orbital overlap,
hydrogen bonding and dispersion interactions are also both affected by
a change in the electron density distribution. The interactions in our
water/\aza\ complexes have contributions from hydrogen bonding,
dispersion interactions and weak orbital mixing. We have seen that the
delicate balance between such interactions is sensitive to the
relative contributions from individual energy terms, prompting the
need for both exact exchange and a good description of the correlation
energy for the reliable prediction of configurations in weakly
interacting systems, such as ours.

\subsection{From benzene to borazine}\label{restwo}
Thus far we have studied the interaction of a water monomer with the
intermediate \aza\ molecule as a model for boron nitride doped
graphene. We have found that an inadequate description of exchange in
xc functionals can alter the binding orientation, despite reasonable
interaction energies being predicted by some of the vdW xc
functionals. Here we establish whether xc functionals without exact
exchange can correctly describe the interaction between the water
monomer and the pure counterparts of \aza: benzene and borazine
(B$_3$N$_3$H$_6$). It is also useful, from a materials design
perspective, to understand how the boron nitride doping affects the
interaction with water, compared to graphene and h-BN; which can be
mimicked to some extent by benzene and borazine,
respectively. Adhering to the use of model systems allows us to
compare interaction energies from DFT and benchmark CCSD(T)
calculations. To this end, we have examined water on benzene and
borazine and we find that in contrast to \aza, these pure systems are
much less challenging, with different functionals able to predict the
same orientation as obtained in previous benchmark
calculations\cite{wb_bind2,wb_bind3,Wu2012}. Previous work has shown
that water tilts towards a carbon atom in benzene, and the complex has
an absolute interaction energy of $-145$ meV calculated with
CCSD(T)\cite{wb_bind2,wb_bind3}.  Similarly, in the water-borazine
complex, water tilts towards a nitrogen atom and the CCSD(T)
interaction energy reported by Wu \etal \cite{Wu2012} is $-92$ meV.

\begin{table}[ht]
\begin{ruledtabular}
\caption{Absolute interaction energies of water to benzene, \aza, and
  borazine using PBE, optB86b-vdW, vdW-DF2 and CCSD(T). The absolute
  interaction energies with DFT correspond to optimized structures,
  and the values for water-\aza\ correspond to the C3 complex.}
\label{purebind}
\begin{threeparttable}
\begin{tabular}{lccc}
Methods     & H$_2$O/Benzene & H$_2$O/\aza & H$_2$O/Borazine \\ \hline
PBE         & $-108$        & $-109$          & $-85$    \\
optB86b-vdW & $-142$        & $-160$          & $-122$   \\
vdW-DF2  & $-137$     & $-155$   & $-129$   \\ 
CCSD(T)     & $-145$\tnote{a}& $-155$\tnote{b}& $-92$\tnote{c,d} \\         
\end{tabular}
\begin{tablenotes}
\item[a]CCSD(T) interaction energy by Min \etal\ with CBS
  extrapolation\cite{wb_bind3}.
\item[b]$\Delta$CCSD(T) interaction energy calculated in Section
  \ref{resone} for the C3 complex.
\item[c] CCSD(T) interaction energy without CBS extrapolation by Wu
  \etal\ \cite{Wu2012}
\item[d]As part of this study our absolute interaction energy is
  $-84\pm4$ meV for water-borazine using DMC.
\end{tablenotes}
\end{threeparttable}
\end{ruledtabular}
\end{table}
We have relaxed several starting geometries of the water-benzene and
water-borazine complexes with PBE and optB86b-vdW and the absolute
interaction energies are shown in Table \ref{purebind}. For the
water-benzene complex, PBE underbinds as previously shown
\cite{wb_bind1}, whilst optB86b-vdW performs very well: only $2\%$
error compared to the benchmark. Moreover with PBE, there is almost no
distinction between water binding to benzene or \aza, but using the
dispersion inclusive optB86b-vdW, the binding on \aza\ is almost 20
meV stronger. Similarly using CCSD(T), our benchmark absolute
interaction energy of the C3 complex ($-155$ meV) is 10 meV stronger
than the absolute interaction energy of water to benzene ($-145$ meV
calculated by Min \etal \cite{wb_bind3}). On borazine the absolute
interaction energy is weaker according to all methods: PBE ($-85$
meV), optB86b-vdW ($-122$ meV) and also with CCSD(T) ($-92$ meV)
\cite{Wu2012}. In all cases, the water tilts towards nitrogen in the
borazine ring\cite{Wu2012} and this tilting, that also occurs on
\aza\ and benzene, is indicative of a weak hydrogen bond forming in
all three complexes. Not only do PBE and optB86b-vdW obtain the
correct binding orientations for water on benzene and borazine, but
also optB86b-vdW yields only a $2$--$3\%$ error for the water-benzene
and C3 complexes. For completeness, we also include the results
obtained with vdW-DF2 in Table \ref{purebind}, which are similar to
those obtained with optB86b-vdW. Using vdW-DF2 the error for
water-benzene is $5\%$ and for the C3 complex the binding energy
agrees with the benchmark. Even so, the wrong orientation is predicted
by PBE, optB86b-vdW and vdW-DF2 for water binding to \aza\ as
discussed in Section \ref{resone}.  Therefore, the prediction of
binding sites with weak interactions on doped surfaces and between
complex organic molecules could be compromised by an inadequate
description of exchange, whilst being correct in pure systems.

Finally, it is interesting to note that water interacts more strongly
with the intermediate \aza\ molecule than either benzene or
borazine. If upon doping graphene with boron and nitrogen a similar
increase in the interaction with water was found then this could
provide a means of tuning the strength of the water substrate
interaction to have specific, targeted wetting properties.

\section{Discussion and Conclusions}\label{conc}
We have calculated three benchmark values of the absolute interaction
energy between a water monomer and 1,2-azaborine using
\ccsdt\ extrapolated to the CBS limit. The lowest energy complex
according to the explicitly correlated, exact exchange methods
(\ccsdt, DMC and MP2) is C3 with an absolute interaction energy of
$-155$ meV from \ccsdt. The DMC energies are, on average, within
$6$--$8\%$ of \ccsdt. Meanwhile, xc functionals including PBE,
dispersion corrected PBE and an array of vdW-DFs, fail to predict the
same binding configuration as \ccsdt, MP2 or DMC. Instead, these xc
functionals indicate that C5 is the lowest energy complex. Although
interatomic many-body dispersion (MBD) forces have previously been
shown to be neglected or wrongly described in DFT xc
functionals\cite{mbddft}, HF and HF-SAPT calculations demonstrate that
in this case, exact exchange corrects the trend without including MBD
forces; thus we can deduce that MBD is not a source of error for the
stability trend. Previous work has shown that for strongly interacting
systems, like CO on Pt(111), exact exchange improves the alignment of
electronic states between the substrate and the adsorbate and to
recover the experimental chemisorption site\cite{Wang2007}. Also, for
defects in semiconductors, exact exchange is a crucial factor in
finding the correct defect states (see e.g. ref
\citenum{DiValentin2006}). Here, it is demonstrated that exact
exchange and the associated changes to the electronic structure plays
a decisive role in the prediction of the lowest energy configuration
for weakly interacting (asymmetric) complexes. As such, it is
imperative that more emphasis is placed on the accurate treatment of
exchange when using DFT to examine weakly bound complexes and
adsorption. The wrong prediction of the configuration for our
relatively small system of water on \aza\ suggests that the
delocalization error is likely to be even more pervasive for larger
systems with more shallow energy minima, such as for physisorption on
doped surfaces and in crystal polymorph prediction.

In terms of the absolute interaction energies, however, the inclusion
of dispersion interactions is essential for weakly bound systems, and
in the systems studied here this is more accurately achieved by
non-local vdW functionals than dispersion corrected PBE, PBE0 and
B3LYP. By comparing absolute interaction energies, we determined that
optB88-vdW and optB86b-vdW are generally the best performing xc
functionals from those tested. Our findings imply that in order to
predict the correct binding configuration as well as the energy, DFT
xc functionals must simultaneously contain exact exchange (so as to
avoid delocalization error) and account for dispersion
interactions. However, even with dispersion corrected hybrid
functionals we were not able to obtain a perfect trend or accurate
interaction energies; echoing the need to develop xc functionals that
contain a better description as well as balance, of both exchange and
non-local correlation\cite{Becke2014}.

Finally, we have found that water binds more strongly to the
intermediate system, \aza\ than to either benzene or borazine. Should
equivalent behavior be observed on doped graphene surfaces, then this
could represent a means of tuning the binding and wetting of
interfacial water to graphene and h-BN. In future work we will focus
on exploiting this possibility.

\acknowledgments We are grateful for support from University College
London and Argonne National Laboratory (ANL) through the Thomas Young
Centre-ANL initiative. Some of the research leading to these results
has received funding from the European Research Council under the
European Union's Seventh Framework Programme (FP/2007-2013) / ERC
Grant Agreement number 616121 (HeteroIce project). A.M. is supported
by the Royal Society through a Wolfson Research Merit
Award. O.A.v.L. acknowledges funding from the Swiss National Science
foundation (No. PPOOP2 138932). This research used resources as part
of an INCITE project (awarded to D.A.)  at the Oak Ridge National
Laboratory, (Titan) which is supported by the Office of Science of the
U.S. Department of Energy (DOE) under Contract
No. DEAC05-00OR22725. This research also used resources of the Argonne
Leadership Computing Facility at Argonne National Laboratory, which is
supported by the Office of Science of the U.S. DOE under contract
DE-AC02-06CH11357. In addition, we are grateful for computing
resources provided by the London Centre for Nanotechnology and
University College London. We would like to thank G. Tocci,
C. Gattinoni, and R. Ramakrishnan for useful discussions.


\begin{thebibliography}{116}%
\makeatletter
\providecommand \@ifxundefined [1]{%
 \@ifx{#1\undefined}
}%
\providecommand \@ifnum [1]{%
 \ifnum #1\expandafter \@firstoftwo
 \else \expandafter \@secondoftwo
 \fi
}%
\providecommand \@ifx [1]{%
 \ifx #1\expandafter \@firstoftwo
 \else \expandafter \@secondoftwo
 \fi
}%
\providecommand \natexlab [1]{#1}%
\providecommand \enquote  [1]{``#1''}%
\providecommand \bibnamefont  [1]{#1}%
\providecommand \bibfnamefont [1]{#1}%
\providecommand \citenamefont [1]{#1}%
\providecommand \href@noop [0]{\@secondoftwo}%
\providecommand \href [0]{\begingroup \@sanitize@url \@href}%
\providecommand \@href[1]{\@@startlink{#1}\@@href}%
\providecommand \@@href[1]{\endgroup#1\@@endlink}%
\providecommand \@sanitize@url [0]{\catcode `\\12\catcode `\$12\catcode
  `\&12\catcode `\#12\catcode `\^12\catcode `\_12\catcode `\%12\relax}%
\providecommand \@@startlink[1]{}%
\providecommand \@@endlink[0]{}%
\providecommand \url  [0]{\begingroup\@sanitize@url \@url }%
\providecommand \@url [1]{\endgroup\@href {#1}{\urlprefix }}%
\providecommand \urlprefix  [0]{URL }%
\providecommand \Eprint [0]{\href }%
\providecommand \doibase [0]{http://dx.doi.org/}%
\providecommand \selectlanguage [0]{\@gobble}%
\providecommand \bibinfo  [0]{\@secondoftwo}%
\providecommand \bibfield  [0]{\@secondoftwo}%
\providecommand \translation [1]{[#1]}%
\providecommand \BibitemOpen [0]{}%
\providecommand \bibitemStop [0]{}%
\providecommand \bibitemNoStop [0]{.\EOS\space}%
\providecommand \EOS [0]{\spacefactor3000\relax}%
\providecommand \BibitemShut  [1]{\csname bibitem#1\endcsname}%
\let\auto@bib@innerbib\@empty
%</preamble>
\bibitem [{\citenamefont {Whalley}(1984)}]{Whalley1984}%
  \BibitemOpen
  \bibfield  {author} {\bibinfo {author} {\bibfnamefont {E.}~\bibnamefont
  {Whalley}},\ }\href@noop {} {\bibfield  {journal} {\bibinfo  {journal} {J.
  Chem. Phys.}\ }\textbf {\bibinfo {volume} {81}},\ \bibinfo {pages} {4087}
  (\bibinfo {year} {1984})}\BibitemShut {NoStop}%
\bibitem [{\citenamefont {Murray}\ and\ \citenamefont
  {Galli}(2012)}]{Galli2012}%
  \BibitemOpen
  \bibfield  {author} {\bibinfo {author} {\bibfnamefont {E.~D.}\ \bibnamefont
  {Murray}}\ and\ \bibinfo {author} {\bibfnamefont {G.}~\bibnamefont {Galli}},\
  }\href@noop {} {\bibfield  {journal} {\bibinfo  {journal} {Phys. Rev. Lett.}\
  }\textbf {\bibinfo {volume} {108}},\ \bibinfo {pages} {105502} (\bibinfo
  {year} {2012})}\BibitemShut {NoStop}%
\bibitem [{\citenamefont {Santra}\ \emph {et~al.}(2011)\citenamefont {Santra},
  \citenamefont {Klime{\v{s}}}, \citenamefont {Alf\`e}, \citenamefont
  {Tkatchenko}, \citenamefont {Slater}, \citenamefont {Michaelides},
  \citenamefont {Car},\ and\ \citenamefont {Scheffler}}]{Santra2011}%
  \BibitemOpen
  \bibfield  {author} {\bibinfo {author} {\bibfnamefont {B.}~\bibnamefont
  {Santra}}, \bibinfo {author} {\bibfnamefont {J.}~\bibnamefont
  {Klime{\v{s}}}}, \bibinfo {author} {\bibfnamefont {D.}~\bibnamefont
  {Alf\`e}}, \bibinfo {author} {\bibfnamefont {A.}~\bibnamefont {Tkatchenko}},
  \bibinfo {author} {\bibfnamefont {B.}~\bibnamefont {Slater}}, \bibinfo
  {author} {\bibfnamefont {A.}~\bibnamefont {Michaelides}}, \bibinfo {author}
  {\bibfnamefont {R.}~\bibnamefont {Car}}, \ and\ \bibinfo {author}
  {\bibfnamefont {M.}~\bibnamefont {Scheffler}},\ }\href@noop {} {\bibfield
  {journal} {\bibinfo  {journal} {Phys. Rev. Lett.}\ }\textbf {\bibinfo
  {volume} {107}},\ \bibinfo {pages} {185701} (\bibinfo {year}
  {2011})}\BibitemShut {NoStop}%
\bibitem [{\citenamefont {Santra}\ \emph {et~al.}(2013)\citenamefont {Santra},
  \citenamefont {Klime\v{s}}, \citenamefont {Tkatchenko}, \citenamefont
  {Alf\`{e}}, \citenamefont {Slater}, \citenamefont {Michaelides},
  \citenamefont {Car},\ and\ \citenamefont {Scheffler}}]{Santra2013}%
  \BibitemOpen
  \bibfield  {author} {\bibinfo {author} {\bibfnamefont {B.}~\bibnamefont
  {Santra}}, \bibinfo {author} {\bibfnamefont {J.}~\bibnamefont {Klime\v{s}}},
  \bibinfo {author} {\bibfnamefont {A.}~\bibnamefont {Tkatchenko}}, \bibinfo
  {author} {\bibfnamefont {D.}~\bibnamefont {Alf\`{e}}}, \bibinfo {author}
  {\bibfnamefont {B.}~\bibnamefont {Slater}}, \bibinfo {author} {\bibfnamefont
  {A.}~\bibnamefont {Michaelides}}, \bibinfo {author} {\bibfnamefont
  {R.}~\bibnamefont {Car}}, \ and\ \bibinfo {author} {\bibfnamefont
  {M.}~\bibnamefont {Scheffler}},\ }\href@noop {} {\bibfield  {journal}
  {\bibinfo  {journal} {J. Chem. Phys.}\ }\textbf {\bibinfo {volume} {139}},\
  \bibinfo {pages} {154702} (\bibinfo {year} {2013})}\BibitemShut {NoStop}%
\bibitem [{\citenamefont {Gillan}\ \emph {et~al.}(2013)\citenamefont {Gillan},
  \citenamefont {Alf\`{e}}, \citenamefont {Bygrave}, \citenamefont {Taylor},\
  and\ \citenamefont {Manby}}]{Gillan2013}%
  \BibitemOpen
  \bibfield  {author} {\bibinfo {author} {\bibfnamefont {M.~J.}\ \bibnamefont
  {Gillan}}, \bibinfo {author} {\bibfnamefont {D.}~\bibnamefont {Alf\`{e}}},
  \bibinfo {author} {\bibfnamefont {P.~J.}\ \bibnamefont {Bygrave}}, \bibinfo
  {author} {\bibfnamefont {C.~R.}\ \bibnamefont {Taylor}}, \ and\ \bibinfo
  {author} {\bibfnamefont {F.~R.}\ \bibnamefont {Manby}},\ }\href@noop {}
  {\bibfield  {journal} {\bibinfo  {journal} {J. Chem. Phys.}\ }\textbf
  {\bibinfo {volume} {139}},\ \bibinfo {pages} {114101} (\bibinfo {year}
  {2013})}\BibitemShut {NoStop}%
\bibitem [{\citenamefont {Slater}\ and\ \citenamefont
  {Quigley}(2014)}]{Slater2014}%
  \BibitemOpen
  \bibfield  {author} {\bibinfo {author} {\bibfnamefont {B.}~\bibnamefont
  {Slater}}\ and\ \bibinfo {author} {\bibfnamefont {D.}~\bibnamefont
  {Quigley}},\ }\href@noop {} {\bibfield  {journal} {\bibinfo  {journal}
  {Nature Mater.}\ }\textbf {\bibinfo {volume} {13}},\ \bibinfo {pages} {670}
  (\bibinfo {year} {2014})}\BibitemShut {NoStop}%
\bibitem [{\citenamefont {Santra}\ \emph {et~al.}(2008)\citenamefont {Santra},
  \citenamefont {Michaelides}, \citenamefont {Fuchs}, \citenamefont
  {Tkatchenko}, \citenamefont {Filippi},\ and\ \citenamefont
  {Scheffler}}]{Santra2008}%
  \BibitemOpen
  \bibfield  {author} {\bibinfo {author} {\bibfnamefont {B.}~\bibnamefont
  {Santra}}, \bibinfo {author} {\bibfnamefont {A.}~\bibnamefont {Michaelides}},
  \bibinfo {author} {\bibfnamefont {M.}~\bibnamefont {Fuchs}}, \bibinfo
  {author} {\bibfnamefont {A.}~\bibnamefont {Tkatchenko}}, \bibinfo {author}
  {\bibfnamefont {C.}~\bibnamefont {Filippi}}, \ and\ \bibinfo {author}
  {\bibfnamefont {M.}~\bibnamefont {Scheffler}},\ }\href@noop {} {\bibfield
  {journal} {\bibinfo  {journal} {J. Chem. Phys.}\ }\textbf {\bibinfo {volume}
  {129}},\ \bibinfo {pages} {194111} (\bibinfo {year} {2008})}\BibitemShut
  {NoStop}%
\bibitem [{\citenamefont {Pedulla}\ \emph {et~al.}(1998)\citenamefont
  {Pedulla}, \citenamefont {Kim},\ and\ \citenamefont {Jordan}}]{Pedulla1998}%
  \BibitemOpen
  \bibfield  {author} {\bibinfo {author} {\bibfnamefont {J.~M.}\ \bibnamefont
  {Pedulla}}, \bibinfo {author} {\bibfnamefont {K.}~\bibnamefont {Kim}}, \ and\
  \bibinfo {author} {\bibfnamefont {K.~D.}\ \bibnamefont {Jordan}},\
  }\href@noop {} {\bibfield  {journal} {\bibinfo  {journal} {Chem. Phys.
  Lett.}\ }\textbf {\bibinfo {volume} {291}},\ \bibinfo {pages} {78} (\bibinfo
  {year} {1998})}\BibitemShut {NoStop}%
\bibitem [{\citenamefont {Price}(2009)}]{Price2009}%
  \BibitemOpen
  \bibfield  {author} {\bibinfo {author} {\bibfnamefont {S.~L.}\ \bibnamefont
  {Price}},\ }\href@noop {} {\bibfield  {journal} {\bibinfo  {journal} {Acc.
  Chem. Res.}\ }\textbf {\bibinfo {volume} {42}},\ \bibinfo {pages} {117}
  (\bibinfo {year} {2009})}\BibitemShut {NoStop}%
\bibitem [{\citenamefont {Price}(2014)}]{Price2014}%
  \BibitemOpen
  \bibfield  {author} {\bibinfo {author} {\bibfnamefont {S.~L.}\ \bibnamefont
  {Price}},\ }\href@noop {} {\bibfield  {journal} {\bibinfo  {journal} {Chem.
  Soc. Rev.}\ }\textbf {\bibinfo {volume} {43}},\ \bibinfo {pages} {2098}
  (\bibinfo {year} {2014})}\BibitemShut {NoStop}%
\bibitem [{\citenamefont {Timasheff}(1993)}]{Timasheff1993}%
  \BibitemOpen
  \bibfield  {author} {\bibinfo {author} {\bibfnamefont {S.~N.}\ \bibnamefont
  {Timasheff}},\ }\href@noop {} {\bibfield  {journal} {\bibinfo  {journal}
  {Annu. Rev. Biophys. Biomol. Struct.}\ }\textbf {\bibinfo {volume} {22}},\
  \bibinfo {pages} {67} (\bibinfo {year} {1993})}\BibitemShut {NoStop}%
\bibitem [{\citenamefont {Siria}\ \emph {et~al.}(2013)\citenamefont {Siria},
  \citenamefont {Poncharal}, \citenamefont {Biance}, \citenamefont {Fulcrand},
  \citenamefont {Blase}, \citenamefont {Purcell},\ and\ \citenamefont
  {Bocquet}}]{Siria}%
  \BibitemOpen
  \bibfield  {author} {\bibinfo {author} {\bibfnamefont {A.}~\bibnamefont
  {Siria}}, \bibinfo {author} {\bibfnamefont {P.}~\bibnamefont {Poncharal}},
  \bibinfo {author} {\bibfnamefont {A.-L.}\ \bibnamefont {Biance}}, \bibinfo
  {author} {\bibfnamefont {R.}~\bibnamefont {Fulcrand}}, \bibinfo {author}
  {\bibfnamefont {X.}~\bibnamefont {Blase}}, \bibinfo {author} {\bibfnamefont
  {S.~T.}\ \bibnamefont {Purcell}}, \ and\ \bibinfo {author} {\bibfnamefont
  {L.}~\bibnamefont {Bocquet}},\ }\href@noop {} {\bibfield  {journal} {\bibinfo
   {journal} {Nature}\ }\textbf {\bibinfo {volume} {494}},\ \bibinfo {pages}
  {455} (\bibinfo {year} {2013})}\BibitemShut {NoStop}%
\bibitem [{\citenamefont {Lei}\ \emph {et~al.}(2013)\citenamefont {Lei},
  \citenamefont {Portehault}, \citenamefont {Liu}, \citenamefont {Qin},\ and\
  \citenamefont {Chen}}]{Lei2013}%
  \BibitemOpen
  \bibfield  {author} {\bibinfo {author} {\bibfnamefont {W.}~\bibnamefont
  {Lei}}, \bibinfo {author} {\bibfnamefont {D.}~\bibnamefont {Portehault}},
  \bibinfo {author} {\bibfnamefont {D.}~\bibnamefont {Liu}}, \bibinfo {author}
  {\bibfnamefont {S.}~\bibnamefont {Qin}}, \ and\ \bibinfo {author}
  {\bibfnamefont {Y.}~\bibnamefont {Chen}},\ }\href@noop {} {\bibfield
  {journal} {\bibinfo  {journal} {Nat. Commun.}\ }\textbf {\bibinfo {volume}
  {4}},\ \bibinfo {pages} {1777} (\bibinfo {year} {2013})}\BibitemShut
  {NoStop}%
\bibitem [{\citenamefont {Pakdel}\ \emph {et~al.}(2011)\citenamefont {Pakdel},
  \citenamefont {Zhi}, \citenamefont {Bando}, \citenamefont {Nakayama},\ and\
  \citenamefont {Golberg}}]{bn_exp3}%
  \BibitemOpen
  \bibfield  {author} {\bibinfo {author} {\bibfnamefont {A.}~\bibnamefont
  {Pakdel}}, \bibinfo {author} {\bibfnamefont {C.}~\bibnamefont {Zhi}},
  \bibinfo {author} {\bibfnamefont {Y.}~\bibnamefont {Bando}}, \bibinfo
  {author} {\bibfnamefont {T.}~\bibnamefont {Nakayama}}, \ and\ \bibinfo
  {author} {\bibfnamefont {D.}~\bibnamefont {Golberg}},\ }\href@noop {}
  {\bibfield  {journal} {\bibinfo  {journal} {ACS Nano}\ }\textbf {\bibinfo
  {volume} {5}},\ \bibinfo {pages} {6507} (\bibinfo {year} {2011})}\BibitemShut
  {NoStop}%
\bibitem [{\citenamefont {Taherian}\ \emph {et~al.}(2013)\citenamefont
  {Taherian}, \citenamefont {Marcon}, \citenamefont {van~der Vegt},\ and\
  \citenamefont {Leroy}}]{graph2}%
  \BibitemOpen
  \bibfield  {author} {\bibinfo {author} {\bibfnamefont {F.}~\bibnamefont
  {Taherian}}, \bibinfo {author} {\bibfnamefont {V.}~\bibnamefont {Marcon}},
  \bibinfo {author} {\bibfnamefont {N.~F.~A.}\ \bibnamefont {van~der Vegt}}, \
  and\ \bibinfo {author} {\bibfnamefont {F.}~\bibnamefont {Leroy}},\
  }\href@noop {} {\bibfield  {journal} {\bibinfo  {journal} {Langmuir}\
  }\textbf {\bibinfo {volume} {29}},\ \bibinfo {pages} {1457} (\bibinfo {year}
  {2013})}\BibitemShut {NoStop}%
\bibitem [{\citenamefont {Shim}\ \emph {et~al.}(2012)\citenamefont {Shim},
  \citenamefont {Lui}, \citenamefont {Ko}, \citenamefont {Yu}, \citenamefont
  {Kim}, \citenamefont {Heinz},\ and\ \citenamefont {Ryu}}]{graph3}%
  \BibitemOpen
  \bibfield  {author} {\bibinfo {author} {\bibfnamefont {J.}~\bibnamefont
  {Shim}}, \bibinfo {author} {\bibfnamefont {C.~H.}\ \bibnamefont {Lui}},
  \bibinfo {author} {\bibfnamefont {T.~Y.}\ \bibnamefont {Ko}}, \bibinfo
  {author} {\bibfnamefont {Y.-J.}\ \bibnamefont {Yu}}, \bibinfo {author}
  {\bibfnamefont {P.}~\bibnamefont {Kim}}, \bibinfo {author} {\bibfnamefont
  {T.~F.}\ \bibnamefont {Heinz}}, \ and\ \bibinfo {author} {\bibfnamefont
  {S.}~\bibnamefont {Ryu}},\ }\href@noop {} {\bibfield  {journal} {\bibinfo
  {journal} {Nano Lett.}\ }\textbf {\bibinfo {volume} {12}},\ \bibinfo {pages}
  {648} (\bibinfo {year} {2012})}\BibitemShut {NoStop}%
\bibitem [{\citenamefont {Ci}\ \emph {et~al.}(2010)\citenamefont {Ci},
  \citenamefont {Song}, \citenamefont {Jin}, \citenamefont {Jariwala},
  \citenamefont {Wu}, \citenamefont {Li}, \citenamefont {Srivastava},
  \citenamefont {Wang}, \citenamefont {Storr}, \citenamefont {Balicas} \emph
  {et~al.}}]{hybBNG}%
  \BibitemOpen
  \bibfield  {author} {\bibinfo {author} {\bibfnamefont {L.}~\bibnamefont
  {Ci}}, \bibinfo {author} {\bibfnamefont {L.}~\bibnamefont {Song}}, \bibinfo
  {author} {\bibfnamefont {C.}~\bibnamefont {Jin}}, \bibinfo {author}
  {\bibfnamefont {D.}~\bibnamefont {Jariwala}}, \bibinfo {author}
  {\bibfnamefont {D.}~\bibnamefont {Wu}}, \bibinfo {author} {\bibfnamefont
  {Y.}~\bibnamefont {Li}}, \bibinfo {author} {\bibfnamefont {A.}~\bibnamefont
  {Srivastava}}, \bibinfo {author} {\bibfnamefont {Z.}~\bibnamefont {Wang}},
  \bibinfo {author} {\bibfnamefont {K.}~\bibnamefont {Storr}}, \bibinfo
  {author} {\bibfnamefont {L.}~\bibnamefont {Balicas}},  \emph {et~al.},\
  }\href@noop {} {\bibfield  {journal} {\bibinfo  {journal} {Nature Mater.}\
  }\textbf {\bibinfo {volume} {9}},\ \bibinfo {pages} {430} (\bibinfo {year}
  {2010})}\BibitemShut {NoStop}%
\bibitem [{\citenamefont {Liu}\ \emph {et~al.}(2013)\citenamefont {Liu},
  \citenamefont {Ma}, \citenamefont {Shi}, \citenamefont {Zhou}, \citenamefont
  {Gong}, \citenamefont {Lei}, \citenamefont {Yang}, \citenamefont {Zhang},
  \citenamefont {Yu}, \citenamefont {Hackenberg} \emph {et~al.}}]{hybBNG1}%
  \BibitemOpen
  \bibfield  {author} {\bibinfo {author} {\bibfnamefont {Z.}~\bibnamefont
  {Liu}}, \bibinfo {author} {\bibfnamefont {L.}~\bibnamefont {Ma}}, \bibinfo
  {author} {\bibfnamefont {G.}~\bibnamefont {Shi}}, \bibinfo {author}
  {\bibfnamefont {W.}~\bibnamefont {Zhou}}, \bibinfo {author} {\bibfnamefont
  {Y.}~\bibnamefont {Gong}}, \bibinfo {author} {\bibfnamefont {S.}~\bibnamefont
  {Lei}}, \bibinfo {author} {\bibfnamefont {X.}~\bibnamefont {Yang}}, \bibinfo
  {author} {\bibfnamefont {J.}~\bibnamefont {Zhang}}, \bibinfo {author}
  {\bibfnamefont {J.}~\bibnamefont {Yu}}, \bibinfo {author} {\bibfnamefont
  {K.~P.}\ \bibnamefont {Hackenberg}},  \emph {et~al.},\ }\href@noop {}
  {\bibfield  {journal} {\bibinfo  {journal} {Nature Nanotech.}\ }\textbf
  {\bibinfo {volume} {8}},\ \bibinfo {pages} {119} (\bibinfo {year}
  {2013})}\BibitemShut {NoStop}%
\bibitem [{\citenamefont {Zheng}\ \emph {et~al.}(2013)\citenamefont {Zheng},
  \citenamefont {Jiao}, \citenamefont {Ge}, \citenamefont {Jaroniec},\ and\
  \citenamefont {Qiao}}]{synBNDG}%
  \BibitemOpen
  \bibfield  {author} {\bibinfo {author} {\bibfnamefont {Y.}~\bibnamefont
  {Zheng}}, \bibinfo {author} {\bibfnamefont {Y.}~\bibnamefont {Jiao}},
  \bibinfo {author} {\bibfnamefont {L.}~\bibnamefont {Ge}}, \bibinfo {author}
  {\bibfnamefont {M.}~\bibnamefont {Jaroniec}}, \ and\ \bibinfo {author}
  {\bibfnamefont {S.~Z.}\ \bibnamefont {Qiao}},\ }\href@noop {} {\bibfield
  {journal} {\bibinfo  {journal} {Angew. Chem.}\ }\textbf {\bibinfo {volume}
  {125}},\ \bibinfo {pages} {3192} (\bibinfo {year} {2013})}\BibitemShut
  {NoStop}%
\bibitem [{\citenamefont {Ding}\ \emph {et~al.}(2011)\citenamefont {Ding},
  \citenamefont {Iannuzzi},\ and\ \citenamefont {Hutter}}]{bn_exp}%
  \BibitemOpen
  \bibfield  {author} {\bibinfo {author} {\bibfnamefont {Y.}~\bibnamefont
  {Ding}}, \bibinfo {author} {\bibfnamefont {M.}~\bibnamefont {Iannuzzi}}, \
  and\ \bibinfo {author} {\bibfnamefont {J.}~\bibnamefont {Hutter}},\
  }\href@noop {} {\bibfield  {journal} {\bibinfo  {journal} {J. Phys. Chem. C}\
  }\textbf {\bibinfo {volume} {115}},\ \bibinfo {pages} {13685} (\bibinfo
  {year} {2011})}\BibitemShut {NoStop}%
\bibitem [{\citenamefont {Boinovich}\ \emph {et~al.}(2012)\citenamefont
  {Boinovich}, \citenamefont {Emelyanenko}, \citenamefont {Pashinin},
  \citenamefont {Lee}, \citenamefont {Drelich},\ and\ \citenamefont
  {Yap}}]{bn_exp4}%
  \BibitemOpen
  \bibfield  {author} {\bibinfo {author} {\bibfnamefont {L.~B.}\ \bibnamefont
  {Boinovich}}, \bibinfo {author} {\bibfnamefont {A.~M.}\ \bibnamefont
  {Emelyanenko}}, \bibinfo {author} {\bibfnamefont {A.~S.}\ \bibnamefont
  {Pashinin}}, \bibinfo {author} {\bibfnamefont {C.~H.}\ \bibnamefont {Lee}},
  \bibinfo {author} {\bibfnamefont {J.}~\bibnamefont {Drelich}}, \ and\
  \bibinfo {author} {\bibfnamefont {Y.~K.}\ \bibnamefont {Yap}},\ }\href@noop
  {} {\bibfield  {journal} {\bibinfo  {journal} {Langmuir}\ }\textbf {\bibinfo
  {volume} {28}},\ \bibinfo {pages} {1206} (\bibinfo {year}
  {2012})}\BibitemShut {NoStop}%
\bibitem [{\citenamefont {Gordillo}\ and\ \citenamefont
  {Mart\'\i}(2011)}]{Marti2011}%
  \BibitemOpen
  \bibfield  {author} {\bibinfo {author} {\bibfnamefont {M.~C.}\ \bibnamefont
  {Gordillo}}\ and\ \bibinfo {author} {\bibfnamefont {J.}~\bibnamefont
  {Mart\'\i}},\ }\href {\doibase 10.1103/PhysRevE.84.011602} {\bibfield
  {journal} {\bibinfo  {journal} {Phys. Rev. E}\ }\textbf {\bibinfo {volume}
  {84}},\ \bibinfo {pages} {011602} (\bibinfo {year} {2011})}\BibitemShut
  {NoStop}%
\bibitem [{\citenamefont {Rafiee}\ \emph {et~al.}(2012)\citenamefont {Rafiee},
  \citenamefont {Mi}, \citenamefont {Gullapalli}, \citenamefont {Thomas},
  \citenamefont {Yavari}, \citenamefont {Shi}, \citenamefont {Ajayan},\ and\
  \citenamefont {Koratkar}}]{graph4}%
  \BibitemOpen
  \bibfield  {author} {\bibinfo {author} {\bibfnamefont {J.}~\bibnamefont
  {Rafiee}}, \bibinfo {author} {\bibfnamefont {X.}~\bibnamefont {Mi}}, \bibinfo
  {author} {\bibfnamefont {H.}~\bibnamefont {Gullapalli}}, \bibinfo {author}
  {\bibfnamefont {A.~V.}\ \bibnamefont {Thomas}}, \bibinfo {author}
  {\bibfnamefont {F.}~\bibnamefont {Yavari}}, \bibinfo {author} {\bibfnamefont
  {Y.}~\bibnamefont {Shi}}, \bibinfo {author} {\bibfnamefont {P.~M.}\
  \bibnamefont {Ajayan}}, \ and\ \bibinfo {author} {\bibfnamefont {N.~A.}\
  \bibnamefont {Koratkar}},\ }\href@noop {} {\bibfield  {journal} {\bibinfo
  {journal} {Nature Mater.}\ }\textbf {\bibinfo {volume} {11}},\ \bibinfo
  {pages} {217} (\bibinfo {year} {2012})}\BibitemShut {NoStop}%
\bibitem [{\citenamefont {Li}\ \emph {et~al.}(2013)\citenamefont {Li},
  \citenamefont {Wang}, \citenamefont {Kozbial}, \citenamefont {Shenoy},
  \citenamefont {Zhou}, \citenamefont {McGinley}, \citenamefont {Ireland},
  \citenamefont {Morganstein}, \citenamefont {Kunkel}, \citenamefont {Surwade}
  \emph {et~al.}}]{graph5}%
  \BibitemOpen
  \bibfield  {author} {\bibinfo {author} {\bibfnamefont {Z.}~\bibnamefont
  {Li}}, \bibinfo {author} {\bibfnamefont {Y.}~\bibnamefont {Wang}}, \bibinfo
  {author} {\bibfnamefont {A.}~\bibnamefont {Kozbial}}, \bibinfo {author}
  {\bibfnamefont {G.}~\bibnamefont {Shenoy}}, \bibinfo {author} {\bibfnamefont
  {F.}~\bibnamefont {Zhou}}, \bibinfo {author} {\bibfnamefont {R.}~\bibnamefont
  {McGinley}}, \bibinfo {author} {\bibfnamefont {P.}~\bibnamefont {Ireland}},
  \bibinfo {author} {\bibfnamefont {B.}~\bibnamefont {Morganstein}}, \bibinfo
  {author} {\bibfnamefont {A.}~\bibnamefont {Kunkel}}, \bibinfo {author}
  {\bibfnamefont {S.~P.}\ \bibnamefont {Surwade}},  \emph {et~al.},\
  }\href@noop {} {\bibfield  {journal} {\bibinfo  {journal} {Nature Mater.}\
  }\textbf {\bibinfo {volume} {12}},\ \bibinfo {pages} {925} (\bibinfo {year}
  {2013})}\BibitemShut {NoStop}%
\bibitem [{\citenamefont {Jenness}\ \emph {et~al.}(2010)\citenamefont
  {Jenness}, \citenamefont {Karalti},\ and\ \citenamefont
  {Jordan}}]{Jenness2010}%
  \BibitemOpen
  \bibfield  {author} {\bibinfo {author} {\bibfnamefont {G.~R.}\ \bibnamefont
  {Jenness}}, \bibinfo {author} {\bibfnamefont {O.}~\bibnamefont {Karalti}}, \
  and\ \bibinfo {author} {\bibfnamefont {K.~D.}\ \bibnamefont {Jordan}},\
  }\href@noop {} {\bibfield  {journal} {\bibinfo  {journal} {Phys. Chem. Chem.
  Phys.}\ }\textbf {\bibinfo {volume} {12}},\ \bibinfo {pages} {6375} (\bibinfo
  {year} {2010})}\BibitemShut {NoStop}%
\bibitem [{\citenamefont {Hamada}(2012)}]{Hamada2012}%
  \BibitemOpen
  \bibfield  {author} {\bibinfo {author} {\bibfnamefont {I.}~\bibnamefont
  {Hamada}},\ }\href {\doibase 10.1103/PhysRevB.86.195436} {\bibfield
  {journal} {\bibinfo  {journal} {Phys. Rev. B}\ }\textbf {\bibinfo {volume}
  {86}},\ \bibinfo {pages} {195436} (\bibinfo {year} {2012})}\BibitemShut
  {NoStop}%
\bibitem [{\citenamefont {Ma}\ \emph {et~al.}(2011)\citenamefont {Ma},
  \citenamefont {Michaelides}, \citenamefont {Alf{\`e}}, \citenamefont
  {Schimka}, \citenamefont {Kresse},\ and\ \citenamefont {Wang}}]{alfe2}%
  \BibitemOpen
  \bibfield  {author} {\bibinfo {author} {\bibfnamefont {J.}~\bibnamefont
  {Ma}}, \bibinfo {author} {\bibfnamefont {A.}~\bibnamefont {Michaelides}},
  \bibinfo {author} {\bibfnamefont {D.}~\bibnamefont {Alf{\`e}}}, \bibinfo
  {author} {\bibfnamefont {L.}~\bibnamefont {Schimka}}, \bibinfo {author}
  {\bibfnamefont {G.}~\bibnamefont {Kresse}}, \ and\ \bibinfo {author}
  {\bibfnamefont {E.}~\bibnamefont {Wang}},\ }\href@noop {} {\bibfield
  {journal} {\bibinfo  {journal} {Phys. Rev. B}\ }\textbf {\bibinfo {volume}
  {84}},\ \bibinfo {pages} {033402} (\bibinfo {year} {2011})}\BibitemShut
  {NoStop}%
\bibitem [{\citenamefont {Voloshina}\ \emph {et~al.}(2011)\citenamefont
  {Voloshina}, \citenamefont {Usvyat}, \citenamefont {Sch{\"u}tz},
  \citenamefont {Dedkov},\ and\ \citenamefont {Paulus}}]{wg_bind1}%
  \BibitemOpen
  \bibfield  {author} {\bibinfo {author} {\bibfnamefont {E.}~\bibnamefont
  {Voloshina}}, \bibinfo {author} {\bibfnamefont {D.}~\bibnamefont {Usvyat}},
  \bibinfo {author} {\bibfnamefont {M.}~\bibnamefont {Sch{\"u}tz}}, \bibinfo
  {author} {\bibfnamefont {Y.}~\bibnamefont {Dedkov}}, \ and\ \bibinfo {author}
  {\bibfnamefont {B.}~\bibnamefont {Paulus}},\ }\href@noop {} {\bibfield
  {journal} {\bibinfo  {journal} {Phys. Chem. Chem. Phys.}\ }\textbf {\bibinfo
  {volume} {13}},\ \bibinfo {pages} {12041} (\bibinfo {year}
  {2011})}\BibitemShut {NoStop}%
\bibitem [{\citenamefont {Rubeš}\ \emph {et~al.}(2009)\citenamefont
  {Rubeš}, \citenamefont {Nachtigall}, \citenamefont {Vondrášek},\ and\
  \citenamefont {Bludský}}]{Rubes2009}%
  \BibitemOpen
  \bibfield  {author} {\bibinfo {author} {\bibfnamefont {M.}~\bibnamefont
  {Rubeš}}, \bibinfo {author} {\bibfnamefont {P.}~\bibnamefont {Nachtigall}},
  \bibinfo {author} {\bibfnamefont {J.}~\bibnamefont {Vondrášek}}, \ and\
  \bibinfo {author} {\bibfnamefont {O.}~\bibnamefont {Bludský}},\ }\href@noop
  {} {\bibfield  {journal} {\bibinfo  {journal} {J. Phys. Chem. C}\ }\textbf
  {\bibinfo {volume} {113}},\ \bibinfo {pages} {8412} (\bibinfo {year}
  {2009})}\BibitemShut {NoStop}%
\bibitem [{\citenamefont {Feller}\ and\ \citenamefont
  {Jordan}(2000)}]{Feller2000}%
  \BibitemOpen
  \bibfield  {author} {\bibinfo {author} {\bibfnamefont {D.}~\bibnamefont
  {Feller}}\ and\ \bibinfo {author} {\bibfnamefont {K.~D.}\ \bibnamefont
  {Jordan}},\ }\href@noop {} {\bibfield  {journal} {\bibinfo  {journal} {J.
  Phys. Chem. A}\ }\textbf {\bibinfo {volume} {104}},\ \bibinfo {pages} {9971}
  (\bibinfo {year} {2000})}\BibitemShut {NoStop}%
\bibitem [{\citenamefont {Xu}\ \emph {et~al.}(2005)\citenamefont {Xu},
  \citenamefont {Irle}, \citenamefont {Musaev},\ and\ \citenamefont
  {Lin}}]{Xu2005}%
  \BibitemOpen
  \bibfield  {author} {\bibinfo {author} {\bibfnamefont {S.}~\bibnamefont
  {Xu}}, \bibinfo {author} {\bibfnamefont {S.}~\bibnamefont {Irle}}, \bibinfo
  {author} {\bibfnamefont {D.~G.}\ \bibnamefont {Musaev}}, \ and\ \bibinfo
  {author} {\bibfnamefont {M.~C.}\ \bibnamefont {Lin}},\ }\href@noop {}
  {\bibfield  {journal} {\bibinfo  {journal} {J. Phys. Chem. A}\ }\textbf
  {\bibinfo {volume} {109}},\ \bibinfo {pages} {9563} (\bibinfo {year}
  {2005})}\BibitemShut {NoStop}%
\bibitem [{\citenamefont {Sudiarta}\ and\ \citenamefont
  {Geldart}(2006)}]{Sudiarta2006}%
  \BibitemOpen
  \bibfield  {author} {\bibinfo {author} {\bibfnamefont {I.~W.}\ \bibnamefont
  {Sudiarta}}\ and\ \bibinfo {author} {\bibfnamefont {D.~J.~W.}\ \bibnamefont
  {Geldart}},\ }\href@noop {} {\bibfield  {journal} {\bibinfo  {journal} {J.
  Phys. Chem. A}\ }\textbf {\bibinfo {volume} {110}},\ \bibinfo {pages} {10501}
  (\bibinfo {year} {2006})}\BibitemShut {NoStop}%
\bibitem [{\citenamefont {Zhao}\ \emph {et~al.}(2005)\citenamefont {Zhao},
  \citenamefont {Tishchenko},\ and\ \citenamefont {Truhlar}}]{wb_bind2}%
  \BibitemOpen
  \bibfield  {author} {\bibinfo {author} {\bibfnamefont {Y.}~\bibnamefont
  {Zhao}}, \bibinfo {author} {\bibfnamefont {O.}~\bibnamefont {Tishchenko}}, \
  and\ \bibinfo {author} {\bibfnamefont {D.~G.}\ \bibnamefont {Truhlar}},\
  }\href@noop {} {\bibfield  {journal} {\bibinfo  {journal} {J. Phys. Chem. B}\
  }\textbf {\bibinfo {volume} {109}},\ \bibinfo {pages} {19046} (\bibinfo
  {year} {2005})}\BibitemShut {NoStop}%
\bibitem [{\citenamefont {Min}\ \emph {et~al.}(2008)\citenamefont {Min},
  \citenamefont {Lee}, \citenamefont {Lee}, \citenamefont {Kim}, \citenamefont
  {Kim},\ and\ \citenamefont {Kim}}]{wb_bind3}%
  \BibitemOpen
  \bibfield  {author} {\bibinfo {author} {\bibfnamefont {S.~K.}\ \bibnamefont
  {Min}}, \bibinfo {author} {\bibfnamefont {E.~C.}\ \bibnamefont {Lee}},
  \bibinfo {author} {\bibfnamefont {H.~M.}\ \bibnamefont {Lee}}, \bibinfo
  {author} {\bibfnamefont {D.~Y.}\ \bibnamefont {Kim}}, \bibinfo {author}
  {\bibfnamefont {D.}~\bibnamefont {Kim}}, \ and\ \bibinfo {author}
  {\bibfnamefont {K.~S.}\ \bibnamefont {Kim}},\ }\href@noop {} {\bibfield
  {journal} {\bibinfo  {journal} {J. Comput. Chem.}\ }\textbf {\bibinfo
  {volume} {29}},\ \bibinfo {pages} {1208} (\bibinfo {year}
  {2008})}\BibitemShut {NoStop}%
\bibitem [{\citenamefont {Feller}(1999)}]{wb_bind1}%
  \BibitemOpen
  \bibfield  {author} {\bibinfo {author} {\bibfnamefont {D.}~\bibnamefont
  {Feller}},\ }\href@noop {} {\bibfield  {journal} {\bibinfo  {journal} {J.
  Phys. Chem. A}\ }\textbf {\bibinfo {volume} {103}},\ \bibinfo {pages} {7558}
  (\bibinfo {year} {1999})}\BibitemShut {NoStop}%
\bibitem [{\citenamefont {Ma}\ \emph {et~al.}(2009)\citenamefont {Ma},
  \citenamefont {Alf{\`e}}, \citenamefont {Michaelides},\ and\ \citenamefont
  {Wang}}]{alfe1}%
  \BibitemOpen
  \bibfield  {author} {\bibinfo {author} {\bibfnamefont {J.}~\bibnamefont
  {Ma}}, \bibinfo {author} {\bibfnamefont {D.}~\bibnamefont {Alf{\`e}}},
  \bibinfo {author} {\bibfnamefont {A.}~\bibnamefont {Michaelides}}, \ and\
  \bibinfo {author} {\bibfnamefont {E.}~\bibnamefont {Wang}},\ }\href@noop {}
  {\bibfield  {journal} {\bibinfo  {journal} {J. Chem. Phys.}\ }\textbf
  {\bibinfo {volume} {130}},\ \bibinfo {pages} {154303} (\bibinfo {year}
  {2009})}\BibitemShut {NoStop}%
\bibitem [{\citenamefont {Wu}\ \emph {et~al.}(2012)\citenamefont {Wu},
  \citenamefont {Yan}, \citenamefont {Chen}, \citenamefont {Dai},\ and\
  \citenamefont {Zhong}}]{Wu2012}%
  \BibitemOpen
  \bibfield  {author} {\bibinfo {author} {\bibfnamefont {J.}~\bibnamefont
  {Wu}}, \bibinfo {author} {\bibfnamefont {H.}~\bibnamefont {Yan}}, \bibinfo
  {author} {\bibfnamefont {H.}~\bibnamefont {Chen}}, \bibinfo {author}
  {\bibfnamefont {G.}~\bibnamefont {Dai}}, \ and\ \bibinfo {author}
  {\bibfnamefont {A.}~\bibnamefont {Zhong}},\ }\href@noop {} {\bibfield
  {journal} {\bibinfo  {journal} {Comput. Theor. Chem.}\ }\textbf {\bibinfo
  {volume} {984}},\ \bibinfo {pages} {51} (\bibinfo {year} {2012})}\BibitemShut
  {NoStop}%
\bibitem [{\citenamefont {Marcon}\ \emph {et~al.}(2007)\citenamefont {Marcon},
  \citenamefont {von Lilienfeld},\ and\ \citenamefont {Andrienko}}]{homoBenz}%
  \BibitemOpen
  \bibfield  {author} {\bibinfo {author} {\bibfnamefont {V.}~\bibnamefont
  {Marcon}}, \bibinfo {author} {\bibfnamefont {O.~A.}\ \bibnamefont {von
  Lilienfeld}}, \ and\ \bibinfo {author} {\bibfnamefont {D.}~\bibnamefont
  {Andrienko}},\ }\href@noop {} {\bibfield  {journal} {\bibinfo  {journal} {J.
  Chem. Phys.}\ }\textbf {\bibinfo {volume} {127}},\ \bibinfo {eid} {064305}
  (\bibinfo {year} {2007})}\BibitemShut {NoStop}%
\bibitem [{\citenamefont {Bj\"{o}rkman}\ \emph {et~al.}(2012)\citenamefont
  {Bj\"{o}rkman}, \citenamefont {Gulans}, \citenamefont {Krasheninnikov},\ and\
  \citenamefont {Nieminen}}]{Bjorkman2012}%
  \BibitemOpen
  \bibfield  {author} {\bibinfo {author} {\bibfnamefont {T.}~\bibnamefont
  {Bj\"{o}rkman}}, \bibinfo {author} {\bibfnamefont {A.}~\bibnamefont
  {Gulans}}, \bibinfo {author} {\bibfnamefont {A.~V.}\ \bibnamefont
  {Krasheninnikov}}, \ and\ \bibinfo {author} {\bibfnamefont {R.~M.}\
  \bibnamefont {Nieminen}},\ }\href@noop {} {\bibfield  {journal} {\bibinfo
  {journal} {Phys. Rev. Lett.}\ }\textbf {\bibinfo {volume} {108}},\ \bibinfo
  {pages} {235502} (\bibinfo {year} {2012})}\BibitemShut {NoStop}%
\bibitem [{\citenamefont {Bj\"{o}rkman}(2014)}]{Bjorkman2014}%
  \BibitemOpen
  \bibfield  {author} {\bibinfo {author} {\bibfnamefont {T.}~\bibnamefont
  {Bj\"{o}rkman}},\ }\href {\doibase http://dx.doi.org/10.1063/1.4893329}
  {\bibfield  {journal} {\bibinfo  {journal} {J. Chem. Phys.}\ }\textbf
  {\bibinfo {volume} {141}},\ \bibinfo {eid} {074708} (\bibinfo {year}
  {2014})}\BibitemShut {NoStop}%
\bibitem [{\citenamefont {Antony}\ and\ \citenamefont
  {Grimme}(2006)}]{Antony2006}%
  \BibitemOpen
  \bibfield  {author} {\bibinfo {author} {\bibfnamefont {J.}~\bibnamefont
  {Antony}}\ and\ \bibinfo {author} {\bibfnamefont {S.}~\bibnamefont
  {Grimme}},\ }\href@noop {} {\bibfield  {journal} {\bibinfo  {journal} {Phys.
  Chem. Chem. Phys.}\ }\textbf {\bibinfo {volume} {8}},\ \bibinfo {pages}
  {5287} (\bibinfo {year} {2006})}\BibitemShut {NoStop}%
\bibitem [{\citenamefont {Graziano}\ \emph {et~al.}(2012)\citenamefont
  {Graziano}, \citenamefont {Klime\v{s}}, \citenamefont {Fernandez-Alonso},\
  and\ \citenamefont {Michaelides}}]{Graziano2012}%
  \BibitemOpen
  \bibfield  {author} {\bibinfo {author} {\bibfnamefont {G.}~\bibnamefont
  {Graziano}}, \bibinfo {author} {\bibfnamefont {J.}~\bibnamefont
  {Klime\v{s}}}, \bibinfo {author} {\bibfnamefont {F.}~\bibnamefont
  {Fernandez-Alonso}}, \ and\ \bibinfo {author} {\bibfnamefont
  {A.}~\bibnamefont {Michaelides}},\ }\href@noop {} {\bibfield  {journal}
  {\bibinfo  {journal} {J. Phys. Condens. Matter}\ }\textbf {\bibinfo {volume}
  {24}},\ \bibinfo {pages} {424216} (\bibinfo {year} {2012})}\BibitemShut
  {NoStop}%
\bibitem [{\citenamefont {Carrasco}\ \emph {et~al.}(2013)\citenamefont
  {Carrasco}, \citenamefont {Klime\v{s}},\ and\ \citenamefont
  {Michaelides}}]{Carrasco2013}%
  \BibitemOpen
  \bibfield  {author} {\bibinfo {author} {\bibfnamefont {J.}~\bibnamefont
  {Carrasco}}, \bibinfo {author} {\bibfnamefont {J.}~\bibnamefont
  {Klime\v{s}}}, \ and\ \bibinfo {author} {\bibfnamefont {A.}~\bibnamefont
  {Michaelides}},\ }\href@noop {} {\bibfield  {journal} {\bibinfo  {journal}
  {J. Chem. Phys.}\ }\textbf {\bibinfo {volume} {138}},\ \bibinfo {pages}
  {024708} (\bibinfo {year} {2013})}\BibitemShut {NoStop}%
\bibitem [{\citenamefont {Thonhauser}\ \emph {et~al.}(2006)\citenamefont
  {Thonhauser}, \citenamefont {Puzder},\ and\ \citenamefont
  {Langreth}}]{Langreth2006}%
  \BibitemOpen
  \bibfield  {author} {\bibinfo {author} {\bibfnamefont {T.}~\bibnamefont
  {Thonhauser}}, \bibinfo {author} {\bibfnamefont {A.}~\bibnamefont {Puzder}},
  \ and\ \bibinfo {author} {\bibfnamefont {D.~C.}\ \bibnamefont {Langreth}},\
  }\href {\doibase http://dx.doi.org/10.1063/1.2189230} {\bibfield  {journal}
  {\bibinfo  {journal} {J. Chem. Phys.}\ }\textbf {\bibinfo {volume} {124}},\
  \bibinfo {eid} {164106} (\bibinfo {year} {2006})}\BibitemShut {NoStop}%
\bibitem [{\citenamefont {Kanai}\ and\ \citenamefont
  {Grossman}(2009)}]{exchangeDFT}%
  \BibitemOpen
  \bibfield  {author} {\bibinfo {author} {\bibfnamefont {Y.}~\bibnamefont
  {Kanai}}\ and\ \bibinfo {author} {\bibfnamefont {J.~C.}\ \bibnamefont
  {Grossman}},\ }\href@noop {} {\bibfield  {journal} {\bibinfo  {journal}
  {Phys. Rev. A}\ }\textbf {\bibinfo {volume} {80}},\ \bibinfo {pages} {032504}
  (\bibinfo {year} {2009})}\BibitemShut {NoStop}%
\bibitem [{\citenamefont {Santra}\ \emph {et~al.}(2009)\citenamefont {Santra},
  \citenamefont {Michaelides},\ and\ \citenamefont {Scheffler}}]{biswajit}%
  \BibitemOpen
  \bibfield  {author} {\bibinfo {author} {\bibfnamefont {B.}~\bibnamefont
  {Santra}}, \bibinfo {author} {\bibfnamefont {A.}~\bibnamefont {Michaelides}},
  \ and\ \bibinfo {author} {\bibfnamefont {M.}~\bibnamefont {Scheffler}},\
  }\href@noop {} {\bibfield  {journal} {\bibinfo  {journal} {J. Chem. Phys.}\
  }\textbf {\bibinfo {volume} {131}},\ \bibinfo {eid} {124509} (\bibinfo {year}
  {2009})}\BibitemShut {NoStop}%
\bibitem [{\citenamefont {Hohenberg}\ and\ \citenamefont {Kohn}(1964)}]{HK}%
  \BibitemOpen
  \bibfield  {author} {\bibinfo {author} {\bibfnamefont {P.}~\bibnamefont
  {Hohenberg}}\ and\ \bibinfo {author} {\bibfnamefont {W.}~\bibnamefont
  {Kohn}},\ }\href@noop {} {\bibfield  {journal} {\bibinfo  {journal} {Phys.
  Rev.}\ }\textbf {\bibinfo {volume} {136}},\ \bibinfo {pages} {B864} (\bibinfo
  {year} {1964})}\BibitemShut {NoStop}%
\bibitem [{\citenamefont {Kohn}\ and\ \citenamefont {Sham}(1965)}]{KS}%
  \BibitemOpen
  \bibfield  {author} {\bibinfo {author} {\bibfnamefont {W.}~\bibnamefont
  {Kohn}}\ and\ \bibinfo {author} {\bibfnamefont {L.~J.}\ \bibnamefont
  {Sham}},\ }\href@noop {} {\bibfield  {journal} {\bibinfo  {journal} {Phys.
  Rev.}\ }\textbf {\bibinfo {volume} {140}},\ \bibinfo {pages} {A1133}
  (\bibinfo {year} {1965})}\BibitemShut {NoStop}%
\bibitem [{\citenamefont {Klime\v{s}}\ and\ \citenamefont
  {Michaelides}(2012)}]{vdwpers}%
  \BibitemOpen
  \bibfield  {author} {\bibinfo {author} {\bibfnamefont {J.}~\bibnamefont
  {Klime\v{s}}}\ and\ \bibinfo {author} {\bibfnamefont {A.}~\bibnamefont
  {Michaelides}},\ }\href@noop {} {\bibfield  {journal} {\bibinfo  {journal}
  {J. Chem. Phys.}\ }\textbf {\bibinfo {volume} {137}},\ \bibinfo {pages}
  {120901} (\bibinfo {year} {2012})}\BibitemShut {NoStop}%
\bibitem [{\citenamefont {Cohen}\ \emph {et~al.}(2008)\citenamefont {Cohen},
  \citenamefont {Mori-S{\'a}nchez},\ and\ \citenamefont {Yang}}]{cohen}%
  \BibitemOpen
  \bibfield  {author} {\bibinfo {author} {\bibfnamefont {A.~J.}\ \bibnamefont
  {Cohen}}, \bibinfo {author} {\bibfnamefont {P.}~\bibnamefont
  {Mori-S{\'a}nchez}}, \ and\ \bibinfo {author} {\bibfnamefont
  {W.}~\bibnamefont {Yang}},\ }\href@noop {} {\bibfield  {journal} {\bibinfo
  {journal} {Science}\ }\textbf {\bibinfo {volume} {321}},\ \bibinfo {pages}
  {792} (\bibinfo {year} {2008})}\BibitemShut {NoStop}%
\bibitem [{\citenamefont {Becke}(2014)}]{Becke2014}%
  \BibitemOpen
  \bibfield  {author} {\bibinfo {author} {\bibfnamefont {A.~D.}\ \bibnamefont
  {Becke}},\ }\href@noop {} {\bibfield  {journal} {\bibinfo  {journal} {J.
  Chem. Phys.}\ }\textbf {\bibinfo {volume} {140}},\ \bibinfo {eid} {18A301}
  (\bibinfo {year} {2014})}\BibitemShut {NoStop}%
\bibitem [{\citenamefont {Burke}(2012)}]{Kieron2012}%
  \BibitemOpen
  \bibfield  {author} {\bibinfo {author} {\bibfnamefont {K.}~\bibnamefont
  {Burke}},\ }\href@noop {} {\bibfield  {journal} {\bibinfo  {journal} {J.
  Chem. Phys.}\ }\textbf {\bibinfo {volume} {136}},\ \bibinfo {pages} {150901}
  (\bibinfo {year} {2012})}\BibitemShut {NoStop}%
\bibitem [{\citenamefont {Shulenburger}\ and\ \citenamefont
  {Mattsson}(2013)}]{Mattsson}%
  \BibitemOpen
  \bibfield  {author} {\bibinfo {author} {\bibfnamefont {L.}~\bibnamefont
  {Shulenburger}}\ and\ \bibinfo {author} {\bibfnamefont {T.~R.}\ \bibnamefont
  {Mattsson}},\ }\href@noop {} {\bibfield  {journal} {\bibinfo  {journal}
  {Phys. Rev. B}\ }\textbf {\bibinfo {volume} {88}},\ \bibinfo {pages} {245117}
  (\bibinfo {year} {2013})}\BibitemShut {NoStop}%
\bibitem [{\citenamefont {Wang}\ \emph {et~al.}(2013)\citenamefont {Wang},
  \citenamefont {Deible},\ and\ \citenamefont {Jordan}}]{Jordan2013}%
  \BibitemOpen
  \bibfield  {author} {\bibinfo {author} {\bibfnamefont {F.-F.}\ \bibnamefont
  {Wang}}, \bibinfo {author} {\bibfnamefont {M.~J.}\ \bibnamefont {Deible}}, \
  and\ \bibinfo {author} {\bibfnamefont {K.~D.}\ \bibnamefont {Jordan}},\
  }\href@noop {} {\bibfield  {journal} {\bibinfo  {journal} {J. Phys. Chem. A}\
  }\textbf {\bibinfo {volume} {117}},\ \bibinfo {pages} {7606} (\bibinfo {year}
  {2013})}\BibitemShut {NoStop}%
\bibitem [{\citenamefont {Dubeck\'{y}}\ \emph {et~al.}(2013)\citenamefont
  {Dubeck\'{y}}, \citenamefont {Jure\v{c}ka}, \citenamefont {Derian},
  \citenamefont {Hobza}, \citenamefont {Otyepka},\ and\ \citenamefont
  {Mitas}}]{Mitas}%
  \BibitemOpen
  \bibfield  {author} {\bibinfo {author} {\bibfnamefont {M.}~\bibnamefont
  {Dubeck\'{y}}}, \bibinfo {author} {\bibfnamefont {P.}~\bibnamefont
  {Jure\v{c}ka}}, \bibinfo {author} {\bibfnamefont {R.}~\bibnamefont {Derian}},
  \bibinfo {author} {\bibfnamefont {P.}~\bibnamefont {Hobza}}, \bibinfo
  {author} {\bibfnamefont {M.}~\bibnamefont {Otyepka}}, \ and\ \bibinfo
  {author} {\bibfnamefont {L.}~\bibnamefont {Mitas}},\ }\href@noop {}
  {\bibfield  {journal} {\bibinfo  {journal} {J. Chem. Theory Comput.}\
  }\textbf {\bibinfo {volume} {9}},\ \bibinfo {pages} {4287} (\bibinfo {year}
  {2013})}\BibitemShut {NoStop}%
\bibitem [{\citenamefont {Dunning~Jr}(1989)}]{dunning1}%
  \BibitemOpen
  \bibfield  {author} {\bibinfo {author} {\bibfnamefont {T.~H.}\ \bibnamefont
  {Dunning~Jr}},\ }\href@noop {} {\bibfield  {journal} {\bibinfo  {journal} {J.
  Chem. Phys.}\ }\textbf {\bibinfo {volume} {90}},\ \bibinfo {pages} {1007}
  (\bibinfo {year} {1989})}\BibitemShut {NoStop}%
\bibitem [{\citenamefont {Kendall}\ \emph {et~al.}(1992)\citenamefont
  {Kendall}, \citenamefont {Dunning~Jr},\ and\ \citenamefont
  {Harrison}}]{dunning2}%
  \BibitemOpen
  \bibfield  {author} {\bibinfo {author} {\bibfnamefont {R.~A.}\ \bibnamefont
  {Kendall}}, \bibinfo {author} {\bibfnamefont {T.~H.}\ \bibnamefont
  {Dunning~Jr}}, \ and\ \bibinfo {author} {\bibfnamefont {R.~J.}\ \bibnamefont
  {Harrison}},\ }\href@noop {} {\bibfield  {journal} {\bibinfo  {journal} {J.
  Chem. Phys.}\ }\textbf {\bibinfo {volume} {96}},\ \bibinfo {pages} {6796}
  (\bibinfo {year} {1992})}\BibitemShut {NoStop}%
\bibitem [{\citenamefont {Woon}\ and\ \citenamefont
  {Dunning~Jr}(1993)}]{dunning3}%
  \BibitemOpen
  \bibfield  {author} {\bibinfo {author} {\bibfnamefont {D.~E.}\ \bibnamefont
  {Woon}}\ and\ \bibinfo {author} {\bibfnamefont {T.~H.}\ \bibnamefont
  {Dunning~Jr}},\ }\href@noop {} {\bibfield  {journal} {\bibinfo  {journal} {J.
  Chem. Phys.}\ }\textbf {\bibinfo {volume} {98}},\ \bibinfo {pages} {1358}
  (\bibinfo {year} {1993})}\BibitemShut {NoStop}%
\bibitem [{Note1()}]{Note1}%
  \BibitemOpen
  \bibinfo {note} {We have also investigated the magnitude of basis set
  superposition error by applying Boys and Bernardi's counterpoise
  correction\protect \cite {cpBB}, but the correction was not included in the
  CBS extrapolation (see SI\cite {SI_ref} for more details).}\BibitemShut
  {Stop}%
\bibitem [{\citenamefont {Burns}\ \emph {et~al.}(2014)\citenamefont {Burns},
  \citenamefont {Marshall},\ and\ \citenamefont {Sherrill}}]{Sherrill}%
  \BibitemOpen
  \bibfield  {author} {\bibinfo {author} {\bibfnamefont {L.~A.}\ \bibnamefont
  {Burns}}, \bibinfo {author} {\bibfnamefont {M.~S.}\ \bibnamefont {Marshall}},
  \ and\ \bibinfo {author} {\bibfnamefont {C.~D.}\ \bibnamefont {Sherrill}},\
  }\href@noop {} {\bibfield  {journal} {\bibinfo  {journal} {J. Chem. Theory
  Comput.}\ }\textbf {\bibinfo {volume} {10}},\ \bibinfo {pages} {49} (\bibinfo
  {year} {2014})}\BibitemShut {NoStop}%
\bibitem [{\citenamefont {Truhlar}(1998)}]{cbs1}%
  \BibitemOpen
  \bibfield  {author} {\bibinfo {author} {\bibfnamefont {D.~G.}\ \bibnamefont
  {Truhlar}},\ }\href@noop {} {\bibfield  {journal} {\bibinfo  {journal} {Chem.
  Phys. Lett.}\ }\textbf {\bibinfo {volume} {294}},\ \bibinfo {pages} {45}
  (\bibinfo {year} {1998})}\BibitemShut {NoStop}%
\bibitem [{\citenamefont {Halkier}\ \emph
  {et~al.}(1999{\natexlab{a}})\citenamefont {Halkier}, \citenamefont {Klopper},
  \citenamefont {Helgaker}, \citenamefont {J{\o}rgensen},\ and\ \citenamefont
  {Taylor}}]{cbs2a}%
  \BibitemOpen
  \bibfield  {author} {\bibinfo {author} {\bibfnamefont {A.}~\bibnamefont
  {Halkier}}, \bibinfo {author} {\bibfnamefont {W.}~\bibnamefont {Klopper}},
  \bibinfo {author} {\bibfnamefont {T.}~\bibnamefont {Helgaker}}, \bibinfo
  {author} {\bibfnamefont {P.}~\bibnamefont {J{\o}rgensen}}, \ and\ \bibinfo
  {author} {\bibfnamefont {P.~R.}\ \bibnamefont {Taylor}},\ }\href@noop {}
  {\bibfield  {journal} {\bibinfo  {journal} {J. Chem. Phys.}\ }\textbf
  {\bibinfo {volume} {111}},\ \bibinfo {pages} {9157} (\bibinfo {year}
  {1999}{\natexlab{a}})}\BibitemShut {NoStop}%
\bibitem [{\citenamefont {Halkier}\ \emph {et~al.}(1998)\citenamefont
  {Halkier}, \citenamefont {Helgaker}, \citenamefont {J{\o}rgensen},
  \citenamefont {Klopper}, \citenamefont {Koch}, \citenamefont {Olsen},\ and\
  \citenamefont {Wilson}}]{cbs2b}%
  \BibitemOpen
  \bibfield  {author} {\bibinfo {author} {\bibfnamefont {A.}~\bibnamefont
  {Halkier}}, \bibinfo {author} {\bibfnamefont {T.}~\bibnamefont {Helgaker}},
  \bibinfo {author} {\bibfnamefont {P.}~\bibnamefont {J{\o}rgensen}}, \bibinfo
  {author} {\bibfnamefont {W.}~\bibnamefont {Klopper}}, \bibinfo {author}
  {\bibfnamefont {H.}~\bibnamefont {Koch}}, \bibinfo {author} {\bibfnamefont
  {J.}~\bibnamefont {Olsen}}, \ and\ \bibinfo {author} {\bibfnamefont {A.~K.}\
  \bibnamefont {Wilson}},\ }\href@noop {} {\bibfield  {journal} {\bibinfo
  {journal} {Chem. Phys. Lett.}\ }\textbf {\bibinfo {volume} {286}},\ \bibinfo
  {pages} {243 } (\bibinfo {year} {1998})}\BibitemShut {NoStop}%
\bibitem [{\citenamefont {Halkier}\ \emph
  {et~al.}(1999{\natexlab{b}})\citenamefont {Halkier}, \citenamefont
  {Helgaker}, \citenamefont {J{\o}rgensen}, \citenamefont {Klopper},\ and\
  \citenamefont {Olsen}}]{cbs3}%
  \BibitemOpen
  \bibfield  {author} {\bibinfo {author} {\bibfnamefont {A.}~\bibnamefont
  {Halkier}}, \bibinfo {author} {\bibfnamefont {T.}~\bibnamefont {Helgaker}},
  \bibinfo {author} {\bibfnamefont {P.}~\bibnamefont {J{\o}rgensen}}, \bibinfo
  {author} {\bibfnamefont {W.}~\bibnamefont {Klopper}}, \ and\ \bibinfo
  {author} {\bibfnamefont {J.}~\bibnamefont {Olsen}},\ }\href@noop {}
  {\bibfield  {journal} {\bibinfo  {journal} {Chem. Phys. Lett.}\ }\textbf
  {\bibinfo {volume} {302}},\ \bibinfo {pages} {437} (\bibinfo {year}
  {1999}{\natexlab{b}})}\BibitemShut {NoStop}%
\bibitem [{\citenamefont {Feller}\ and\ \citenamefont {Peterson}(1999)}]{cbs4}%
  \BibitemOpen
  \bibfield  {author} {\bibinfo {author} {\bibfnamefont {D.}~\bibnamefont
  {Feller}}\ and\ \bibinfo {author} {\bibfnamefont {K.~A.}\ \bibnamefont
  {Peterson}},\ }\href@noop {} {\bibfield  {journal} {\bibinfo  {journal} {J.
  Chem. Phys.}\ }\textbf {\bibinfo {volume} {110}},\ \bibinfo {pages} {8384}
  (\bibinfo {year} {1999})}\BibitemShut {NoStop}%
\bibitem [{\citenamefont {Frisch}\ \emph {et~al.}(2004)\citenamefont {Frisch},
  \citenamefont {Trucks}, \citenamefont {Schlegel}, \citenamefont {Scuseria},
  \citenamefont {Robb}, \citenamefont {Cheeseman}, \citenamefont {Montgomery}
  \emph {et~al.}}]{g03}%
  \BibitemOpen
  \bibfield  {author} {\bibinfo {author} {\bibfnamefont {M.~J.}\ \bibnamefont
  {Frisch}}, \bibinfo {author} {\bibfnamefont {G.~W.}\ \bibnamefont {Trucks}},
  \bibinfo {author} {\bibfnamefont {H.~B.}\ \bibnamefont {Schlegel}}, \bibinfo
  {author} {\bibfnamefont {G.~E.}\ \bibnamefont {Scuseria}}, \bibinfo {author}
  {\bibfnamefont {M.~A.}\ \bibnamefont {Robb}}, \bibinfo {author}
  {\bibfnamefont {J.~R.}\ \bibnamefont {Cheeseman}}, \bibinfo {author}
  {\bibfnamefont {J.~A.}\ \bibnamefont {Montgomery}, \bibfnamefont {Jr.}},
  \emph {et~al.},\ }\href@noop {} {\enquote {\bibinfo {title} {Gaussian 03},}\
  }\bibinfo {howpublished} {Revision D.02, Gaussian, Inc., Wallingford, CT}
  (\bibinfo {year} {2004})\BibitemShut {NoStop}%
\bibitem [{\citenamefont {Giannozzi}\ \emph {et~al.}(2009)\citenamefont
  {Giannozzi}, \citenamefont {Baroni}, \citenamefont {Bonini}, \citenamefont
  {Calandra}, \citenamefont {Car}, \citenamefont {Cavazzoni}, \citenamefont
  {Ceresoli} \emph {et~al.}}]{pwscf}%
  \BibitemOpen
  \bibfield  {author} {\bibinfo {author} {\bibfnamefont {P.}~\bibnamefont
  {Giannozzi}}, \bibinfo {author} {\bibfnamefont {S.}~\bibnamefont {Baroni}},
  \bibinfo {author} {\bibfnamefont {N.}~\bibnamefont {Bonini}}, \bibinfo
  {author} {\bibfnamefont {M.}~\bibnamefont {Calandra}}, \bibinfo {author}
  {\bibfnamefont {R.}~\bibnamefont {Car}}, \bibinfo {author} {\bibfnamefont
  {C.}~\bibnamefont {Cavazzoni}}, \bibinfo {author} {\bibfnamefont
  {D.}~\bibnamefont {Ceresoli}},  \emph {et~al.},\ }\href@noop {} {\bibfield
  {journal} {\bibinfo  {journal} {J. Phys.: Condens. Matter}\ }\textbf
  {\bibinfo {volume} {21}},\ \bibinfo {pages} {395502} (\bibinfo {year}
  {2009})}\BibitemShut {NoStop}%
\bibitem [{\citenamefont {Trail}\ and\ \citenamefont
  {Needs}(2005{\natexlab{a}})}]{TN1}%
  \BibitemOpen
  \bibfield  {author} {\bibinfo {author} {\bibfnamefont {J.}~\bibnamefont
  {Trail}}\ and\ \bibinfo {author} {\bibfnamefont {R.}~\bibnamefont {Needs}},\
  }\href@noop {} {\bibfield  {journal} {\bibinfo  {journal} {J. Chem. Phys.}\
  }\textbf {\bibinfo {volume} {122}},\ \bibinfo {pages} {174109} (\bibinfo
  {year} {2005}{\natexlab{a}})}\BibitemShut {NoStop}%
\bibitem [{\citenamefont {Trail}\ and\ \citenamefont
  {Needs}(2005{\natexlab{b}})}]{TN2}%
  \BibitemOpen
  \bibfield  {author} {\bibinfo {author} {\bibfnamefont {J.}~\bibnamefont
  {Trail}}\ and\ \bibinfo {author} {\bibfnamefont {R.}~\bibnamefont {Needs}},\
  }\href@noop {} {\bibfield  {journal} {\bibinfo  {journal} {J. Chem. Phys.}\
  }\textbf {\bibinfo {volume} {122}},\ \bibinfo {pages} {014112} (\bibinfo
  {year} {2005}{\natexlab{b}})}\BibitemShut {NoStop}%
\bibitem [{\citenamefont {Perdew}\ and\ \citenamefont {Zunger}(1981)}]{LDA}%
  \BibitemOpen
  \bibfield  {author} {\bibinfo {author} {\bibfnamefont {J.~P.}\ \bibnamefont
  {Perdew}}\ and\ \bibinfo {author} {\bibfnamefont {A.}~\bibnamefont
  {Zunger}},\ }\href@noop {} {\bibfield  {journal} {\bibinfo  {journal} {Phys.
  Rev. B}\ }\textbf {\bibinfo {volume} {23}},\ \bibinfo {pages} {5048}
  (\bibinfo {year} {1981})}\BibitemShut {NoStop}%
\bibitem [{\citenamefont {Perdew}\ \emph
  {et~al.}(1996{\natexlab{a}})\citenamefont {Perdew}, \citenamefont {Burke},\
  and\ \citenamefont {Ernzerhof}}]{PBE}%
  \BibitemOpen
  \bibfield  {author} {\bibinfo {author} {\bibfnamefont {J.~P.}\ \bibnamefont
  {Perdew}}, \bibinfo {author} {\bibfnamefont {K.}~\bibnamefont {Burke}}, \
  and\ \bibinfo {author} {\bibfnamefont {M.}~\bibnamefont {Ernzerhof}},\
  }\href@noop {} {\bibfield  {journal} {\bibinfo  {journal} {Phys. Rev. Lett.}\
  }\textbf {\bibinfo {volume} {77}},\ \bibinfo {pages} {3865} (\bibinfo {year}
  {1996}{\natexlab{a}})}\BibitemShut {NoStop}%
\bibitem [{\citenamefont {Alf{\`e}}\ and\ \citenamefont
  {Gillan}(2004)}]{bsplines}%
  \BibitemOpen
  \bibfield  {author} {\bibinfo {author} {\bibfnamefont {D.}~\bibnamefont
  {Alf{\`e}}}\ and\ \bibinfo {author} {\bibfnamefont {M.}~\bibnamefont
  {Gillan}},\ }\href@noop {} {\bibfield  {journal} {\bibinfo  {journal} {Phys.
  Rev. B}\ }\textbf {\bibinfo {volume} {70}} (\bibinfo {year}
  {2004})}\BibitemShut {NoStop}%
\bibitem [{\citenamefont {Needs}\ \emph {et~al.}(2010)\citenamefont {Needs},
  \citenamefont {Towler}, \citenamefont {Drummond},\ and\ \citenamefont
  {R{\'\i}os}}]{casino}%
  \BibitemOpen
  \bibfield  {author} {\bibinfo {author} {\bibfnamefont {R.}~\bibnamefont
  {Needs}}, \bibinfo {author} {\bibfnamefont {M.}~\bibnamefont {Towler}},
  \bibinfo {author} {\bibfnamefont {N.}~\bibnamefont {Drummond}}, \ and\
  \bibinfo {author} {\bibfnamefont {P.~L.}\ \bibnamefont {R{\'\i}os}},\
  }\href@noop {} {\enquote {\bibinfo {title} {Casino version 2.13},}\ }
  (\bibinfo {year} {2010})\BibitemShut {NoStop}%
\bibitem [{\citenamefont {Mitas}\ \emph {et~al.}(1991)\citenamefont {Mitas},
  \citenamefont {Shirley},\ and\ \citenamefont {Ceperley}}]{locapp}%
  \BibitemOpen
  \bibfield  {author} {\bibinfo {author} {\bibfnamefont {L.}~\bibnamefont
  {Mitas}}, \bibinfo {author} {\bibfnamefont {E.~L.}\ \bibnamefont {Shirley}},
  \ and\ \bibinfo {author} {\bibfnamefont {D.}~\bibnamefont {Ceperley}},\
  }\href@noop {} {\bibfield  {journal} {\bibinfo  {journal} {J. Chem. Phys.}\
  }\textbf {\bibinfo {volume} {95}} (\bibinfo {year} {1991})}\BibitemShut
  {NoStop}%
\bibitem [{\citenamefont {Kresse}\ and\ \citenamefont {Hafner}(1993)}]{vasp1}%
  \BibitemOpen
  \bibfield  {author} {\bibinfo {author} {\bibfnamefont {G.}~\bibnamefont
  {Kresse}}\ and\ \bibinfo {author} {\bibfnamefont {J.}~\bibnamefont
  {Hafner}},\ }\href@noop {} {\bibfield  {journal} {\bibinfo  {journal} {Phys.
  Rev. B}\ }\textbf {\bibinfo {volume} {47}},\ \bibinfo {pages} {558} (\bibinfo
  {year} {1993})}\BibitemShut {NoStop}%
\bibitem [{\citenamefont {Kresse}\ and\ \citenamefont {Hafner}(1994)}]{vasp2}%
  \BibitemOpen
  \bibfield  {author} {\bibinfo {author} {\bibfnamefont {G.}~\bibnamefont
  {Kresse}}\ and\ \bibinfo {author} {\bibfnamefont {J.}~\bibnamefont
  {Hafner}},\ }\href@noop {} {\bibfield  {journal} {\bibinfo  {journal} {Phys.
  Rev. B}\ }\textbf {\bibinfo {volume} {49}},\ \bibinfo {pages} {14251}
  (\bibinfo {year} {1994})}\BibitemShut {NoStop}%
\bibitem [{\citenamefont {Kresse}\ and\ \citenamefont
  {Furthm{\"u}ller}(1996{\natexlab{a}})}]{vasp3}%
  \BibitemOpen
  \bibfield  {author} {\bibinfo {author} {\bibfnamefont {G.}~\bibnamefont
  {Kresse}}\ and\ \bibinfo {author} {\bibfnamefont {J.}~\bibnamefont
  {Furthm{\"u}ller}},\ }\href@noop {} {\bibfield  {journal} {\bibinfo
  {journal} {{Comp. Mater. Sci.}}\ }\textbf {\bibinfo {volume} {{6}}},\
  \bibinfo {pages} {15} (\bibinfo {year} {{1996}}{\natexlab{a}})}\BibitemShut
  {NoStop}%
\bibitem [{\citenamefont {Kresse}\ and\ \citenamefont
  {Furthm{\"u}ller}(1996{\natexlab{b}})}]{vasp4}%
  \BibitemOpen
  \bibfield  {author} {\bibinfo {author} {\bibfnamefont {G.}~\bibnamefont
  {Kresse}}\ and\ \bibinfo {author} {\bibfnamefont {J.}~\bibnamefont
  {Furthm{\"u}ller}},\ }\href@noop {} {\bibfield  {journal} {\bibinfo
  {journal} {Phys. Rev. B}\ }\textbf {\bibinfo {volume} {54}},\ \bibinfo
  {pages} {11169} (\bibinfo {year} {1996}{\natexlab{b}})}\BibitemShut {NoStop}%
\bibitem [{\citenamefont {Bl{\"o}chl}(1994)}]{PAW}%
  \BibitemOpen
  \bibfield  {author} {\bibinfo {author} {\bibfnamefont {P.~E.}\ \bibnamefont
  {Bl{\"o}chl}},\ }\href@noop {} {\bibfield  {journal} {\bibinfo  {journal}
  {Phys. Rev. B}\ }\textbf {\bibinfo {volume} {50}},\ \bibinfo {pages} {17953}
  (\bibinfo {year} {1994})}\BibitemShut {NoStop}%
\bibitem [{\citenamefont {Kresse}\ and\ \citenamefont {Joubert}(1999)}]{paw2}%
  \BibitemOpen
  \bibfield  {author} {\bibinfo {author} {\bibfnamefont {G.}~\bibnamefont
  {Kresse}}\ and\ \bibinfo {author} {\bibfnamefont {D.}~\bibnamefont
  {Joubert}},\ }\href@noop {} {\bibfield  {journal} {\bibinfo  {journal} {Phys.
  Rev. B}\ }\textbf {\bibinfo {volume} {59}},\ \bibinfo {pages} {1758}
  (\bibinfo {year} {1999})}\BibitemShut {NoStop}%
\bibitem [{\citenamefont {Adamo}\ and\ \citenamefont {Barone}(1999)}]{PBE0a}%
  \BibitemOpen
  \bibfield  {author} {\bibinfo {author} {\bibfnamefont {C.}~\bibnamefont
  {Adamo}}\ and\ \bibinfo {author} {\bibfnamefont {V.}~\bibnamefont {Barone}},\
  }\href@noop {} {\bibfield  {journal} {\bibinfo  {journal} {J. Chem. Phys.}\
  }\textbf {\bibinfo {volume} {110}},\ \bibinfo {pages} {6158} (\bibinfo {year}
  {1999})}\BibitemShut {NoStop}%
\bibitem [{\citenamefont {Perdew}\ \emph
  {et~al.}(1996{\natexlab{b}})\citenamefont {Perdew}, \citenamefont
  {Ernzerhof},\ and\ \citenamefont {Burke}}]{PBE0b}%
  \BibitemOpen
  \bibfield  {author} {\bibinfo {author} {\bibfnamefont {J.~P.}\ \bibnamefont
  {Perdew}}, \bibinfo {author} {\bibfnamefont {M.}~\bibnamefont {Ernzerhof}}, \
  and\ \bibinfo {author} {\bibfnamefont {K.}~\bibnamefont {Burke}},\
  }\href@noop {} {\bibfield  {journal} {\bibinfo  {journal} {J. Chem. Phys.}\
  }\textbf {\bibinfo {volume} {105}},\ \bibinfo {pages} {9982} (\bibinfo {year}
  {1996}{\natexlab{b}})}\BibitemShut {NoStop}%
\bibitem [{\citenamefont {Becke}(1993)}]{b3lypA}%
  \BibitemOpen
  \bibfield  {author} {\bibinfo {author} {\bibfnamefont {A.~D.}\ \bibnamefont
  {Becke}},\ }\href@noop {} {\bibfield  {journal} {\bibinfo  {journal} {J.
  Chem. Phys.}\ }\textbf {\bibinfo {volume} {98}},\ \bibinfo {pages} {5648}
  (\bibinfo {year} {1993})}\BibitemShut {NoStop}%
\bibitem [{\citenamefont {Lee}\ \emph {et~al.}(1988)\citenamefont {Lee},
  \citenamefont {Yang},\ and\ \citenamefont {Parr}}]{b3lypB}%
  \BibitemOpen
  \bibfield  {author} {\bibinfo {author} {\bibfnamefont {C.}~\bibnamefont
  {Lee}}, \bibinfo {author} {\bibfnamefont {W.}~\bibnamefont {Yang}}, \ and\
  \bibinfo {author} {\bibfnamefont {R.~G.}\ \bibnamefont {Parr}},\ }\href@noop
  {} {\bibfield  {journal} {\bibinfo  {journal} {Phys. Rev. B}\ }\textbf
  {\bibinfo {volume} {37}},\ \bibinfo {pages} {785} (\bibinfo {year}
  {1988})}\BibitemShut {NoStop}%
\bibitem [{\citenamefont {Vosko}\ \emph {et~al.}(1980)\citenamefont {Vosko},
  \citenamefont {Wilk},\ and\ \citenamefont {Nusair}}]{b3lypC}%
  \BibitemOpen
  \bibfield  {author} {\bibinfo {author} {\bibfnamefont {S.~H.}\ \bibnamefont
  {Vosko}}, \bibinfo {author} {\bibfnamefont {L.}~\bibnamefont {Wilk}}, \ and\
  \bibinfo {author} {\bibfnamefont {M.}~\bibnamefont {Nusair}},\ }\href@noop {}
  {\bibfield  {journal} {\bibinfo  {journal} {Can. J. Phys.}\ }\textbf
  {\bibinfo {volume} {58}},\ \bibinfo {pages} {1200} (\bibinfo {year}
  {1980})}\BibitemShut {NoStop}%
\bibitem [{\citenamefont {Stephens}\ \emph {et~al.}(1994)\citenamefont
  {Stephens}, \citenamefont {Devlin}, \citenamefont {Chabalowski},\ and\
  \citenamefont {Frisch}}]{b3lypD}%
  \BibitemOpen
  \bibfield  {author} {\bibinfo {author} {\bibfnamefont {P.~J.}\ \bibnamefont
  {Stephens}}, \bibinfo {author} {\bibfnamefont {F.~J.}\ \bibnamefont
  {Devlin}}, \bibinfo {author} {\bibfnamefont {C.~F.}\ \bibnamefont
  {Chabalowski}}, \ and\ \bibinfo {author} {\bibfnamefont {M.~J.}\ \bibnamefont
  {Frisch}},\ }\href@noop {} {\bibfield  {journal} {\bibinfo  {journal} {J.
  Phys. Chem.}\ }\textbf {\bibinfo {volume} {98}},\ \bibinfo {pages} {11623}
  (\bibinfo {year} {1994})}\BibitemShut {NoStop}%
\bibitem [{\citenamefont {Grimme}(2011)}]{vdwreview}%
  \BibitemOpen
  \bibfield  {author} {\bibinfo {author} {\bibfnamefont {S.}~\bibnamefont
  {Grimme}},\ }\href@noop {} {\bibfield  {journal} {\bibinfo  {journal} {WIREs
  Comput. Mol. Sci.}\ }\textbf {\bibinfo {volume} {1}},\ \bibinfo {pages} {211}
  (\bibinfo {year} {2011})}\BibitemShut {NoStop}%
\bibitem [{\citenamefont {Grimme}(2006)}]{D2}%
  \BibitemOpen
  \bibfield  {author} {\bibinfo {author} {\bibfnamefont {S.}~\bibnamefont
  {Grimme}},\ }\href@noop {} {\bibfield  {journal} {\bibinfo  {journal} {J.
  Comp. Chem.}\ }\textbf {\bibinfo {volume} {27}},\ \bibinfo {pages} {1787}
  (\bibinfo {year} {2006})}\BibitemShut {NoStop}%
\bibitem [{\citenamefont {Tkatchenko}\ and\ \citenamefont
  {Scheffler}(2009)}]{TS}%
  \BibitemOpen
  \bibfield  {author} {\bibinfo {author} {\bibfnamefont {A.}~\bibnamefont
  {Tkatchenko}}\ and\ \bibinfo {author} {\bibfnamefont {M.}~\bibnamefont
  {Scheffler}},\ }\href@noop {} {\bibfield  {journal} {\bibinfo  {journal}
  {Phys. Rev. Lett.}\ }\textbf {\bibinfo {volume} {102}},\ \bibinfo {pages}
  {073005} (\bibinfo {year} {2009})}\BibitemShut {NoStop}%
\bibitem [{\citenamefont {Tkatchenko}\ \emph {et~al.}(2012)\citenamefont
  {Tkatchenko}, \citenamefont {DiStasio}, \citenamefont {Car},\ and\
  \citenamefont {Scheffler}}]{MBD}%
  \BibitemOpen
  \bibfield  {author} {\bibinfo {author} {\bibfnamefont {A.}~\bibnamefont
  {Tkatchenko}}, \bibinfo {author} {\bibfnamefont {R.~A.}\ \bibnamefont
  {DiStasio}}, \bibinfo {author} {\bibfnamefont {R.}~\bibnamefont {Car}}, \
  and\ \bibinfo {author} {\bibfnamefont {M.}~\bibnamefont {Scheffler}},\
  }\href@noop {} {\bibfield  {journal} {\bibinfo  {journal} {Phys. Rev. Lett.}\
  }\textbf {\bibinfo {volume} {108}},\ \bibinfo {pages} {236402} (\bibinfo
  {year} {2012})}\BibitemShut {NoStop}%
\bibitem [{Note2()}]{Note2}%
  \BibitemOpen
  \bibinfo {note} {As the TS and TS+SCS schemes are implemented in the later
  versions of VASP, VASP.5.3.3 was used for these particular
  calculations.}\BibitemShut {Stop}%
\bibitem [{\citenamefont {Dion}\ \emph {et~al.}(2004)\citenamefont {Dion},
  \citenamefont {Rydberg}, \citenamefont {Schr{\"o}der}, \citenamefont
  {Langreth},\ and\ \citenamefont {Lundqvist}}]{vdwDF}%
  \BibitemOpen
  \bibfield  {author} {\bibinfo {author} {\bibfnamefont {M.}~\bibnamefont
  {Dion}}, \bibinfo {author} {\bibfnamefont {H.}~\bibnamefont {Rydberg}},
  \bibinfo {author} {\bibfnamefont {E.}~\bibnamefont {Schr{\"o}der}}, \bibinfo
  {author} {\bibfnamefont {D.~C.}\ \bibnamefont {Langreth}}, \ and\ \bibinfo
  {author} {\bibfnamefont {B.~I.}\ \bibnamefont {Lundqvist}},\ }\href@noop {}
  {\bibfield  {journal} {\bibinfo  {journal} {Phys. Rev. Lett.}\ }\textbf
  {\bibinfo {volume} {92}},\ \bibinfo {pages} {246401} (\bibinfo {year}
  {2004})}\BibitemShut {NoStop}%
\bibitem [{\citenamefont {Klime{\v{s}}}\ \emph {et~al.}(2010)\citenamefont
  {Klime{\v{s}}}, \citenamefont {Bowler},\ and\ \citenamefont
  {Michaelides}}]{vdwfuncs}%
  \BibitemOpen
  \bibfield  {author} {\bibinfo {author} {\bibfnamefont {J.}~\bibnamefont
  {Klime{\v{s}}}}, \bibinfo {author} {\bibfnamefont {D.~R.}\ \bibnamefont
  {Bowler}}, \ and\ \bibinfo {author} {\bibfnamefont {A.}~\bibnamefont
  {Michaelides}},\ }\href@noop {} {\bibfield  {journal} {\bibinfo  {journal}
  {J. Phys.: Condens. Matter}\ }\textbf {\bibinfo {volume} {22}},\ \bibinfo
  {pages} {022201} (\bibinfo {year} {2010})}\BibitemShut {NoStop}%
\bibitem [{\citenamefont {Klime{\v{s}}}\ \emph {et~al.}(2011)\citenamefont
  {Klime{\v{s}}}, \citenamefont {Bowler},\ and\ \citenamefont
  {Michaelides}}]{vdwimp}%
  \BibitemOpen
  \bibfield  {author} {\bibinfo {author} {\bibfnamefont {J.}~\bibnamefont
  {Klime{\v{s}}}}, \bibinfo {author} {\bibfnamefont {D.~R.}\ \bibnamefont
  {Bowler}}, \ and\ \bibinfo {author} {\bibfnamefont {A.}~\bibnamefont
  {Michaelides}},\ }\href@noop {} {\bibfield  {journal} {\bibinfo  {journal}
  {Phys. Rev. B}\ }\textbf {\bibinfo {volume} {83}},\ \bibinfo {pages} {195131}
  (\bibinfo {year} {2011})}\BibitemShut {NoStop}%
\bibitem [{\citenamefont {Lee}\ \emph {et~al.}(2010)\citenamefont {Lee},
  \citenamefont {Murray}, \citenamefont {Kong}, \citenamefont {Lundqvist},\
  and\ \citenamefont {Langreth}}]{vdwDF2}%
  \BibitemOpen
  \bibfield  {author} {\bibinfo {author} {\bibfnamefont {K.}~\bibnamefont
  {Lee}}, \bibinfo {author} {\bibfnamefont {{\'E}.~D.}\ \bibnamefont {Murray}},
  \bibinfo {author} {\bibfnamefont {L.}~\bibnamefont {Kong}}, \bibinfo {author}
  {\bibfnamefont {B.~I.}\ \bibnamefont {Lundqvist}}, \ and\ \bibinfo {author}
  {\bibfnamefont {D.~C.}\ \bibnamefont {Langreth}},\ }\href@noop {} {\bibfield
  {journal} {\bibinfo  {journal} {Phys. Rev. B}\ }\textbf {\bibinfo {volume}
  {82}},\ \bibinfo {pages} {081101} (\bibinfo {year} {2010})}\BibitemShut
  {NoStop}%
\bibitem [{\citenamefont {Cooper}(2010)}]{Cooper2010}%
  \BibitemOpen
  \bibfield  {author} {\bibinfo {author} {\bibfnamefont {V.~R.}\ \bibnamefont
  {Cooper}},\ }\href {\doibase 10.1103/PhysRevB.81.161104} {\bibfield
  {journal} {\bibinfo  {journal} {Phys. Rev. B}\ }\textbf {\bibinfo {volume}
  {81}},\ \bibinfo {pages} {161104} (\bibinfo {year} {2010})}\BibitemShut
  {NoStop}%
\bibitem [{\citenamefont {Berland}\ and\ \citenamefont
  {Hyldgaard}(2014)}]{Hyldgaard2014}%
  \BibitemOpen
  \bibfield  {author} {\bibinfo {author} {\bibfnamefont {K.}~\bibnamefont
  {Berland}}\ and\ \bibinfo {author} {\bibfnamefont {P.}~\bibnamefont
  {Hyldgaard}},\ }\href {\doibase 10.1103/PhysRevB.89.035412} {\bibfield
  {journal} {\bibinfo  {journal} {Phys. Rev. B}\ }\textbf {\bibinfo {volume}
  {89}},\ \bibinfo {pages} {035412} (\bibinfo {year} {2014})}\BibitemShut
  {NoStop}%
\bibitem [{\citenamefont {Hamada}(2014)}]{Hamada2014}%
  \BibitemOpen
  \bibfield  {author} {\bibinfo {author} {\bibfnamefont {I.}~\bibnamefont
  {Hamada}},\ }\href {\doibase 10.1103/PhysRevB.89.121103} {\bibfield
  {journal} {\bibinfo  {journal} {Phys. Rev. B}\ }\textbf {\bibinfo {volume}
  {89}},\ \bibinfo {pages} {121103} (\bibinfo {year} {2014})}\BibitemShut
  {NoStop}%
\bibitem [{\citenamefont {Vydrov}\ and\ \citenamefont
  {Van~Voorhis}(2009)}]{Voorhis2009}%
  \BibitemOpen
  \bibfield  {author} {\bibinfo {author} {\bibfnamefont {O.~A.}\ \bibnamefont
  {Vydrov}}\ and\ \bibinfo {author} {\bibfnamefont {T.}~\bibnamefont
  {Van~Voorhis}},\ }\href {\doibase 10.1103/PhysRevLett.103.063004} {\bibfield
  {journal} {\bibinfo  {journal} {Phys. Rev. Lett.}\ }\textbf {\bibinfo
  {volume} {103}},\ \bibinfo {pages} {063004} (\bibinfo {year}
  {2009})}\BibitemShut {NoStop}%
\bibitem [{\citenamefont {Vydrov}\ and\ \citenamefont
  {Van~Voorhis}(2010)}]{Voorhis2010}%
  \BibitemOpen
  \bibfield  {author} {\bibinfo {author} {\bibfnamefont {O.~A.}\ \bibnamefont
  {Vydrov}}\ and\ \bibinfo {author} {\bibfnamefont {T.}~\bibnamefont
  {Van~Voorhis}},\ }\href {\doibase http://dx.doi.org/10.1063/1.3398840}
  {\bibfield  {journal} {\bibinfo  {journal} {J. Chem. Phys.}\ }\textbf
  {\bibinfo {volume} {132}},\ \bibinfo {eid} {164113} (\bibinfo {year}
  {2010})}\BibitemShut {NoStop}%
\bibitem [{\citenamefont {Sabatini}\ \emph {et~al.}(2013)\citenamefont
  {Sabatini}, \citenamefont {Gorni},\ and\ \citenamefont
  {de~Gironcoli}}]{Gironcoli2013}%
  \BibitemOpen
  \bibfield  {author} {\bibinfo {author} {\bibfnamefont {R.}~\bibnamefont
  {Sabatini}}, \bibinfo {author} {\bibfnamefont {T.}~\bibnamefont {Gorni}}, \
  and\ \bibinfo {author} {\bibfnamefont {S.}~\bibnamefont {de~Gironcoli}},\
  }\href {\doibase 10.1103/PhysRevB.87.041108} {\bibfield  {journal} {\bibinfo
  {journal} {Phys. Rev. B}\ }\textbf {\bibinfo {volume} {87}},\ \bibinfo
  {pages} {041108} (\bibinfo {year} {2013})}\BibitemShut {NoStop}%
\bibitem [{\citenamefont {Wellendorff}\ \emph {et~al.}(2012)\citenamefont
  {Wellendorff}, \citenamefont {Lundgaard}, \citenamefont {M\o{}gelh\o{}j},
  \citenamefont {Petzold}, \citenamefont {Landis}, \citenamefont {N\o{}rskov},
  \citenamefont {Bligaard},\ and\ \citenamefont {Jacobsen}}]{Jacobsen2012}%
  \BibitemOpen
  \bibfield  {author} {\bibinfo {author} {\bibfnamefont {J.}~\bibnamefont
  {Wellendorff}}, \bibinfo {author} {\bibfnamefont {K.~T.}\ \bibnamefont
  {Lundgaard}}, \bibinfo {author} {\bibfnamefont {A.}~\bibnamefont
  {M\o{}gelh\o{}j}}, \bibinfo {author} {\bibfnamefont {V.}~\bibnamefont
  {Petzold}}, \bibinfo {author} {\bibfnamefont {D.~D.}\ \bibnamefont {Landis}},
  \bibinfo {author} {\bibfnamefont {J.~K.}\ \bibnamefont {N\o{}rskov}},
  \bibinfo {author} {\bibfnamefont {T.}~\bibnamefont {Bligaard}}, \ and\
  \bibinfo {author} {\bibfnamefont {K.~W.}\ \bibnamefont {Jacobsen}},\ }\href
  {\doibase 10.1103/PhysRevB.85.235149} {\bibfield  {journal} {\bibinfo
  {journal} {Phys. Rev. B}\ }\textbf {\bibinfo {volume} {85}},\ \bibinfo
  {pages} {235149} (\bibinfo {year} {2012})}\BibitemShut {NoStop}%
\bibitem [{\citenamefont {Bj\"orkman}(2012)}]{BjorkmanPRB2012}%
  \BibitemOpen
  \bibfield  {author} {\bibinfo {author} {\bibfnamefont {T.}~\bibnamefont
  {Bj\"orkman}},\ }\href {\doibase 10.1103/PhysRevB.86.165109} {\bibfield
  {journal} {\bibinfo  {journal} {Phys. Rev. B}\ }\textbf {\bibinfo {volume}
  {86}},\ \bibinfo {pages} {165109} (\bibinfo {year} {2012})}\BibitemShut
  {NoStop}%
\bibitem [{Note3()}]{Note3}%
  \BibitemOpen
  \bibinfo {note} {A 10 \r A\ long cubic cell was used with standard PBE PAW
  potentials and a 500 eV cut-off energy. Convergence criteria of $10^{-6}$ eV
  for the wavefunction optimization and $0.01$ eV/\r A\ for the forces were
  used.}\BibitemShut {Stop}%
\bibitem [{\citenamefont {Liu}\ \emph {et~al.}(2012)\citenamefont {Liu},
  \citenamefont {Pilania}, \citenamefont {Wang},\ and\ \citenamefont
  {Ramprasad}}]{Ramprasad2012}%
  \BibitemOpen
  \bibfield  {author} {\bibinfo {author} {\bibfnamefont {C.-S.}\ \bibnamefont
  {Liu}}, \bibinfo {author} {\bibfnamefont {G.}~\bibnamefont {Pilania}},
  \bibinfo {author} {\bibfnamefont {C.}~\bibnamefont {Wang}}, \ and\ \bibinfo
  {author} {\bibfnamefont {R.}~\bibnamefont {Ramprasad}},\ }\href {\doibase
  10.1021/jp3005844} {\bibfield  {journal} {\bibinfo  {journal} {J. Phys. Chem.
  A}\ }\textbf {\bibinfo {volume} {116}},\ \bibinfo {pages} {9347} (\bibinfo
  {year} {2012})}\BibitemShut {NoStop}%
\bibitem [{\citenamefont {Bu\v{c}ko}\ \emph {et~al.}(2010)\citenamefont
  {Bu\v{c}ko}, \citenamefont {Hafner}, \citenamefont {Leb\`{e}gue},\ and\
  \citenamefont {Angy{\'a}n}}]{Angyan2010}%
  \BibitemOpen
  \bibfield  {author} {\bibinfo {author} {\bibfnamefont {T.}~\bibnamefont
  {Bu\v{c}ko}}, \bibinfo {author} {\bibfnamefont {J.}~\bibnamefont {Hafner}},
  \bibinfo {author} {\bibfnamefont {S.}~\bibnamefont {Leb\`{e}gue}}, \ and\
  \bibinfo {author} {\bibfnamefont {J.~G.}\ \bibnamefont {Angy{\'a}n}},\ }\href
  {\doibase 10.1021/jp106469x} {\bibfield  {journal} {\bibinfo  {journal} {J.
  Phys. Chem. A}\ }\textbf {\bibinfo {volume} {114}},\ \bibinfo {pages} {11814}
  (\bibinfo {year} {2010})}\BibitemShut {NoStop}%
\bibitem [{\citenamefont {Tunega}\ \emph {et~al.}(2012)\citenamefont {Tunega},
  \citenamefont {Bučko},\ and\ \citenamefont {Zaoui}}]{Zaoui2012}%
  \BibitemOpen
  \bibfield  {author} {\bibinfo {author} {\bibfnamefont {D.}~\bibnamefont
  {Tunega}}, \bibinfo {author} {\bibfnamefont {T.}~\bibnamefont {Bučko}}, \
  and\ \bibinfo {author} {\bibfnamefont {A.}~\bibnamefont {Zaoui}},\ }\href
  {\doibase http://dx.doi.org/10.1063/1.4752196} {\bibfield  {journal}
  {\bibinfo  {journal} {J. Chem. Phys.}\ }\textbf {\bibinfo {volume} {137}},\
  \bibinfo {eid} {114105} (\bibinfo {year} {2012})}\BibitemShut {NoStop}%
\bibitem [{SI_()}]{SI_ref}%
  \BibitemOpen
  \href@noop {} {}\bibinfo {note} {See supplemental material at [URL will be
  inserted by AIP] for geometrical details of the binding configurations,
  interaction energies from optimised structures, molecular orbital diagrams,
  and a list of interaction energies from the quantum chemical
  calculations.}\BibitemShut {Stop}%
\bibitem [{\citenamefont {Benali}\ \emph {et~al.}(2014)\citenamefont {Benali},
  \citenamefont {Shulenburger}, \citenamefont {Romero}, \citenamefont {Kim},\
  and\ \citenamefont {von Lilienfeld}}]{Benali2014}%
  \BibitemOpen
  \bibfield  {author} {\bibinfo {author} {\bibfnamefont {A.}~\bibnamefont
  {Benali}}, \bibinfo {author} {\bibfnamefont {L.}~\bibnamefont
  {Shulenburger}}, \bibinfo {author} {\bibfnamefont {N.~A.}\ \bibnamefont
  {Romero}}, \bibinfo {author} {\bibfnamefont {J.}~\bibnamefont {Kim}}, \ and\
  \bibinfo {author} {\bibfnamefont {O.~A.}\ \bibnamefont {von Lilienfeld}},\
  }\href@noop {} {\bibfield  {journal} {\bibinfo  {journal} {J. Chem. Theory
  Comput.}\ }\textbf {\bibinfo {volume} {10}} (\bibinfo {year}
  {2014})}\BibitemShut {NoStop}%
\bibitem [{Note4()}]{Note4}%
  \BibitemOpen
  \bibinfo {note} {The LDA TWs give rise to total energies that are only $\sim
  20$--$30$ meV lower than total energies obtained from PBE TWs, whereby the
  total energies are in the region of $\sim 1500$ eV.}\BibitemShut {Stop}%
\bibitem [{Note5()}]{Note5}%
  \BibitemOpen
  \bibinfo {note} {HF-SAPT calculations were performed using Molpro
  2010\protect \cite {MOLPRO_brief} and an aug-cc-pVDZ basis set for the C3 and
  C5 complexes.}\BibitemShut {Stop}%
\bibitem [{\citenamefont {von Lilienfeld}\ and\ \citenamefont
  {Tkatchenko}(2010)}]{mbddft}%
  \BibitemOpen
  \bibfield  {author} {\bibinfo {author} {\bibfnamefont {O.~A.}\ \bibnamefont
  {von Lilienfeld}}\ and\ \bibinfo {author} {\bibfnamefont {A.}~\bibnamefont
  {Tkatchenko}},\ }\href@noop {} {\bibfield  {journal} {\bibinfo  {journal} {J.
  Chem. Phys.}\ }\textbf {\bibinfo {volume} {132}},\ \bibinfo {eid} {234109}
  (\bibinfo {year} {2010})}\BibitemShut {NoStop}%
\bibitem [{\citenamefont {Wang}\ \emph {et~al.}(2007)\citenamefont {Wang},
  \citenamefont {de~Gironcoli}, \citenamefont {Hush},\ and\ \citenamefont
  {Reimers}}]{Wang2007}%
  \BibitemOpen
  \bibfield  {author} {\bibinfo {author} {\bibfnamefont {Y.}~\bibnamefont
  {Wang}}, \bibinfo {author} {\bibfnamefont {S.}~\bibnamefont {de~Gironcoli}},
  \bibinfo {author} {\bibfnamefont {N.~S.}\ \bibnamefont {Hush}}, \ and\
  \bibinfo {author} {\bibfnamefont {J.~R.}\ \bibnamefont {Reimers}},\
  }\href@noop {} {\bibfield  {journal} {\bibinfo  {journal} {J. Am. Chem.
  Soc.}\ }\textbf {\bibinfo {volume} {129}},\ \bibinfo {pages} {10402}
  (\bibinfo {year} {2007})}\BibitemShut {NoStop}%
\bibitem [{\citenamefont {{Di Valentin}}\ \emph {et~al.}(2006)\citenamefont
  {{Di Valentin}}, \citenamefont {Pacchioni},\ and\ \citenamefont
  {Selloni}}]{DiValentin2006}%
  \BibitemOpen
  \bibfield  {author} {\bibinfo {author} {\bibfnamefont {C.}~\bibnamefont {{Di
  Valentin}}}, \bibinfo {author} {\bibfnamefont {G.}~\bibnamefont {Pacchioni}},
  \ and\ \bibinfo {author} {\bibfnamefont {A.}~\bibnamefont {Selloni}},\
  }\href@noop {} {\bibfield  {journal} {\bibinfo  {journal} {Phys. Rev. Lett.}\
  }\textbf {\bibinfo {volume} {97}},\ \bibinfo {pages} {166803} (\bibinfo
  {year} {2006})}\BibitemShut {NoStop}%
\bibitem [{\citenamefont {Boys}\ and\ \citenamefont {Bernardi}(1970)}]{cpBB}%
  \BibitemOpen
  \bibfield  {author} {\bibinfo {author} {\bibfnamefont {S.~F.}\ \bibnamefont
  {Boys}}\ and\ \bibinfo {author} {\bibfnamefont {F.}~\bibnamefont
  {Bernardi}},\ }\href@noop {} {\bibfield  {journal} {\bibinfo  {journal} {Mol.
  Phys.}\ }\textbf {\bibinfo {volume} {19}},\ \bibinfo {pages} {553} (\bibinfo
  {year} {1970})}\BibitemShut {NoStop}%
\bibitem [{\citenamefont {Werner}\ \emph {et~al.}(2010)\citenamefont {Werner},
  \citenamefont {Knowles}, \citenamefont {Knizia}, \citenamefont {Manby},
  \citenamefont {{Sch\"{u}tz}}, \citenamefont {Celani}, \citenamefont {Korona}
  \emph {et~al.}}]{MOLPRO_brief}%
  \BibitemOpen
  \bibfield  {author} {\bibinfo {author} {\bibfnamefont {H.-J.}\ \bibnamefont
  {Werner}}, \bibinfo {author} {\bibfnamefont {P.~J.}\ \bibnamefont {Knowles}},
  \bibinfo {author} {\bibfnamefont {G.}~\bibnamefont {Knizia}}, \bibinfo
  {author} {\bibfnamefont {F.~R.}\ \bibnamefont {Manby}}, \bibinfo {author}
  {\bibfnamefont {M.}~\bibnamefont {{Sch\"{u}tz}}}, \bibinfo {author}
  {\bibfnamefont {P.}~\bibnamefont {Celani}}, \bibinfo {author} {\bibfnamefont
  {T.}~\bibnamefont {Korona}},  \emph {et~al.},\ }\href@noop {} {\enquote
  {\bibinfo {title} {Molpro, version 2010.1, a package of ab initio
  programs},}\ } (\bibinfo {year} {2010})\BibitemShut {NoStop}%
\end{thebibliography}
\end{document}